\documentclass[twocolumn]{aastex62}
\usepackage{amsmath}
\usepackage{amsfonts}

\usepackage[]{color}

\begin{document}

\title{Force-free wave interaction in magnetar magnetospheres: Computational modeling in axisymmetry}
\shorttitle{Force-free wave interaction in magnetar magnetospheres}
\shortauthors{Mahlmann, Aloy, and Li}

\correspondingauthor{Jens F. Mahlmann}
\email{jens.mahlmann@columbia.edu}

\correspondingauthor{Xinyu Li}
\email{xinyuli@mail.tsinghua.edu.cn}

\author[0000-0002-5349-7116]{Jens F. Mahlmann}
\affiliation{Department of Astronomy and Columbia Astrophysics Laboratory, Columbia University, New York, NY 10027, USA}

\author[0000-0002-5552-7681]{Miguel Á. Aloy}
\affiliation{Department of Astronomy and Astrophysics, Universitat de València, 46100 Burjassot, Spain}
\affiliation{Observatori Astronòmic, Universitat de València, 46980 Paterna, Spain}

\author[0000-0003-0750-3543]{Xinyu Li}
\affiliation{Department of Astronomy, Tsinghua University, Beijing, 100084, China}

\begin{abstract}
Crustal quakes of highly magnetized neutron stars can disrupt their magnetospheres, triggering energetic phenomena like X-ray and fast radio bursts (FRBs). Understanding plasma wave dynamics in these extreme environments is vital for predicting energy transport across scales to the radiation length. This study models relativistic plasma wave interaction in magnetar magnetospheres with force-free electrodynamics simulations. For propagation along curved magnetic field lines, we observe the continuous conversion of Alfvén waves to fast magnetosonic (FMS) waves. The conversion efficiency can be up to three times higher when counter-propagating Alfvén waves interact in the equatorial region. Alfvén waves generate FMS waves of twice their frequency during their first crossing of the magnetosphere. After the initial transient burst of FMS waves, Alfvén waves convert to FMS waves periodically, generating variations on timescales of the magnetospheric Alfvén wave crossing time. This decaying FMS wave tail carries a significant portion (half) of the total energy emitted. Plastic damping of `bouncing' Alfvén waves by the magnetar crust has minimal impact on the FMS efficiency. We discuss the implications of the identified wave phenomena for magnetar observations. Outgoing FMS waves can develop electric zones, potential sources of coherent radiation. Long-wavelength FMS waves could generate FRBs through reconnection beyond the light cylinder.
\end{abstract}


\section{Introduction}

Magnetars are neutron stars with ultrastrong magnetic fields of $10^{14}-10^{16}\,{\rm G}$ \citep[see reviews by][]{Kaspi_2017ARAaA..55..261, Esposito2020}. 
They exhibit rich radiative activities. 
Transient X-ray events include powerful giant flares with luminosities up to $10^{47}\,{\rm erg/s}$ on millisecond timescales, and lower energy outbursts of $10^{35}\,{\rm erg/s}$ with month-to-year long decays. Quasiperiodic oscillations (QPOs) with frequencies ranging from approximately $150\,{\rm Hz}$ to $500\,{\rm Hz}$ are commonly observed
during giant flares \citep{Barat_1983A&A...126..400,Strohmayer_Watts_2005ApJ...632L.111, Israel_2005ApJ...628L..53, Castro-Tirado_2021Natur.600..621}. At least in some cases, these QPOs have been attributed to global torsional vibrations \citep[e.g.][]{Strohmayer_Watts_2006ApJ...653..593,Levin2007,Gabler2011}.
The energy source of these events is likely the magnetic field. 
Sudden global magnetospheric rearrangement could power fast giant flares \citep{Thompson_Duncan_1995MNRAS.275..255,thompson_duncan2001, Parfrey2013,Carrasco:2019aas,Mahlmann2022}, while slower dissipation of magnetospheric twist was proposed to explain persistent magnetar emissions \citep{Beloborodov2013}. 
In addition, the galactic magnetar SGR 1935+2154 is a source of fast radio bursts \citep[FRBs;][]{Chime2020}. FRB polarization and timing information offers insights into the highly magnetized environments of compact astrophysical objects. 
Plasma instabilities in the magnetized environment of magnetars offer a suitable energy reservoir to power the coherent generation of FRBs \citep[see reviews, e.g., in][]{Lyubarsky2021,Petroff2022,Zhang2023}.
How magnetic energy is dissipated in the magnetosphere to power X-ray bursts and FRBs is the key theoretical question that remains to be answered.

One possible mechanism of energy dissipation in the highly relativistic magnetar magnetosphere is nonlinear wave interactions \citep[e.g.,][]{Lyubarsky2018,TenBarge2021,Ripperda2021,Golbraikh2023}.
Starquakes \citep{Blaes1989,thompson_duncan2001} and crustal failures in the stellar interior \citep{li2016,thompson_crust_2017} can perturb the magnetosphere by displacing footpoints of magnetic field lines and excite waves that carry magnetic energy into the extended magnetosphere. Waves in highly magnetized plasmas can be modeled well by force-free electrodynamics (FFE), assuming a vanishing Lorentz force with plasma particles streaming along magnetic field lines \citep{uchida1997,gruzinov1999,Komissarov2002}.
FFE supports two wave modes as solutions to the full nonlinear 
FFE equations: Alfvén waves and fast magnetosonic (FMS) waves. Alfvén waves transport energy along the direction of the background field while FMS waves propagate similarly to waves in vacuum electrodynamics -- across magnetic field lines. \citet{thompson_duncan2001} and \citet{Thompson2022} proposed that the nonlinear interaction between Alfvén waves can drive turbulent cascades to small scales that heat the magnetospheric pair plasma to form a radiatively efficient `fireball' that explains magnetar X-ray bursts.

The theory of wave interactions and their resulting spectra is long debated. \citet{gs1994} studied weak Alfvén turbulence in nonrelativistic incompressible fluids and found leading order four-wave interactions resulting in an anisotropic spectrum.
Three-wave interactions are kinetically prohibited \citep[only possible if one of the involved Alfvén waves is nonpropagating, see][]{ng1996,gs1997}. FMS waves can be excited to enable three-wave interactions in compressible fluids and play an important role in the energy transfer \citep{thompson1998,takamoto2016,takamoto2017}.
Similar conclusions are also found in relativistic turbulence \citep{thompson1998,cho2005,zrake2012,zrake2016,takamoto2016,takamoto2017,Li2019,TenBarge2021,Ripperda2021}.
Numerical models \citep{Li2019,Mahlmann2020b,Ripperda2021,Nattila2022} of relativistic Alfvén turbulence suggest that the turbulent dissipation may be slow compared to competing processes like Alfvén wave absorption by the stellar surface \citep{Li2015}. 
However, these works study Alfvén wave interactions in local settings with uniform background magnetic fields. 
Alfvén wave dynamics in dipole magnetospheres tied to a stellar surface require more complex models.

Recently, nonlinear wave dynamics have been playing a
prominent role in various FRB models \citep{lyubarsky2020,yuan2020,lu_frb_2020}.
Alfvén waves increase in amplitudes compared to the decaying magnetospheric fields as they propagate to the outer magnetosphere.
When they become nonlinear, they open up the magnetosphere to form plasmoids, ignite magnetic reconnection, and drive relativistic blast waves \citep{yuan2020} that could eventually excite radio emission by the synchrotron maser instability \citep[e.g.,][]{Ghisellini2016,Plotnikov2019,Metzger2019}. 
Alfvén waves propagating along curved magnetic fields are a source of FMS waves due to nonlinear coupling with the background magnetic field \citep{yuan2021,chen2024,Bernardi2024}. Long-wavelength FMS pulses are relevant for FRB models because they can perturb the current sheet beyond the Y-point or in the magnetar wind to generate radio emission \citep{lyubarsky2020,Mahlmann2023,wang2023}. In this context, it is crucial to disentangle all the different processes involved in the Alfvén wave dynamics: wave interactions,  cascading of energy from large scales to small scales,
mode conversion, and dissipation.

In this work, we study essential aspects of Alfvén wave interactions in axisymmetric dipolar magnetospheres to derive limits for FMS wave generation enhanced by nonlinear interactions. 
This paper is organized as follows. 
Section~\ref{sec:prelims} reviews the basic theory of FFE magnetospheric modeling and the possible wave interactions (Section~\ref{sec:wavedynamics}). 
We revisit the mode conversion at curved field lines (Section~\ref{sec:modeconvert}) and through nonlinear interactions (Section~\ref{sec:nonlinear}). 
Section~\ref{sec:simulations} describes our basic numerical setup (Section~\ref{sec:setup}) and analyzes various FFE simulations: 
Section~\ref{sec:trains} evaluates the conversion of long wave packets of Alfvén waves to FMS waves in the inner magnetosphere. Section~\ref{sec:enhancement} shows how Alfvén wave collisions
enhance FMS wave generation, and Section~\ref{sec:counterprop} analyzes the long-term dynamics of counter-propagating Alfvén waves. We discuss the implications of our results in Section~\ref{sec:discussion} and summarize our conclusions in Section~\ref{sec:conclusion}.

\section{Theoretical background}
\label{sec:prelims}

\begin{figure}
\centering  \includegraphics[width=0.975\linewidth]{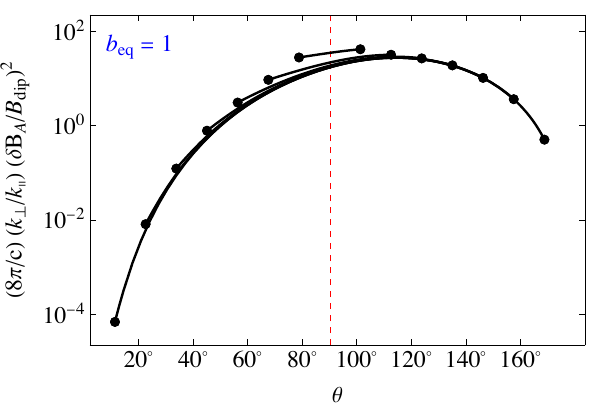}
  \caption{Approximate scaling of the coupling strength between Alfvén waves propagating along curved field lines and the FMS wave electric field (Equation~\ref{eq:couplingestimate}; $b_{\rm eq}=1$). Black dots indicate colatitudes at which field lines are anchored to the star (red line: equator). Alfvén waves injected in the northern hemisphere propagate along dipolar field lines through the magnetosphere (varying their $\theta$ position).}
  \label{fig:COUPLING}
\end{figure}

Deformations of the magnetar crust can generate Alfvén waves propagating along magnetic field lines with wavelengths comparable to a significant fraction of the stellar radius $R_\ast$ \citep[e.g.,][]{Thompson2017}. On such scales, magnetic field dynamics of the highly relativistic magnetar magnetospheres can be described by FFE \citep[see][]{Blandford2002,gruzinov1999, Spitkovsky2006}, the high-magnetization limit of relativistic ideal magnetohydrodynamics (MHD). In this limit, electric ($\mathbf{E}$) and magnetic ($\mathbf{B}$) fields follow from Maxwell's equations:
\begin{align}
    c\nabla\times\mathbf{E}&=-\partial_t\mathbf{B}\label{eq:MAXI}\,,\\
    \nabla\cdot\mathbf{E}&=4\pi\rho\,,\\
    c\nabla\times\mathbf{B}&=4\pi\mathbf{J}+\partial_t\mathbf{E}\,,\label{eq:MaxIII} \\
    \nabla\cdot\mathbf{B}&=0\,.\label{eq:MAXIV}
\end{align}
Here, $\rho$ is a charge density, and $c$ denotes the speed of light. The vanishing Lorentz force constraint $\rho\mathbf{E}+\mathbf{J}/c\times\mathbf{B}=\mathbf{0}$ determines currents ($\mathbf{J}$) that can be written as: 
\begin{align}
    \mathbf{J}_{\rm FFE}=\rho \frac{\mathbf{E}\times\mathbf{B}}{B^2}+\frac{\mathbf{B}\cdot(\nabla\times\mathbf{B})-\mathbf{E}\cdot(\nabla\times\mathbf{E})}{B^2}\mathbf{B}.\label{eq:JFFE}
\end{align}
Equations~(\ref{eq:MAXI}) to (\ref{eq:JFFE}) allow for two wave modes: FMS waves with $\omega=\pm ck$, and Alfvén waves with $\omega=\pm ck_\parallel$. Here, $\omega$ denotes the wave frequency, and $\mathbf{k}$ is the wavevector with norm $k$. Vectors can be decomposed into projections parallel ($\parallel$) and perpendicular ($\perp$) to the magnetic field. Detailed reviews of FFE waves were presented, for example, by \citet{Punsly2003} and \citet{Li2019}. The following sections recap the properties as well as linear and nonlinear wave dynamics of force-free dipolar magnetar magnetospheres. 

\subsection{Waves in the dipole magnetosphere}
\label{sec:wavedynamics}

We consider a dipolar magnetic field in spherical coordinates $(r,\theta,\phi)$ and an orthogonal (but not unitary) basis:
\begin{align}
\begin{split}
 \mathbf{B}_{\rm dip}&=\mu\left(\frac{2\cos\theta}{r^3},\frac{\sin\theta}{r^4},0\right),\\
 B_{\rm dip}&=\frac{\mu}{r^3}\sqrt{3\cos^2\theta+1}.
\end{split}
 \label{eq:dipoleinit}
\end{align}
Here, $\mu=B_* R^3_*/2$ is the magnetic moment of the star, and $B_*$ is the surface magnetic field strength. Deformations or faults in a magnetar crust can generate both Alfvén waves and FMS waves \citep[see][]{Blaes1989,thompson_duncan2001,li2016,thompson_crust_2017}. The high incompressibility of the neutron star matter and crust implies that coupling seismic waves to Alfvén waves is more likely than launching (compressive) FMS waves. Any FMS wave launched from the stellar surface moves across magnetic field lines and carries energy away from the star. This work investigates how much of the energy of Alfvén waves, produced at the footpoints of magnetic field lines and propagating along magnetic flux tubes in the inner magnetosphere, can escape the closed magnetosphere as secondary FMS waves. Displacing magnetic field lines in the toroidal direction with an amplitude $\epsilon_0$ at the stellar surface induces Alfvén wave magnetic fields of relative strength $b=\delta B/B_{\rm dip}$. At the stellar surface, we find\footnote{In axisymmetric dipole magnetospheres, Alfvén waves have a purely toroidal magnetic field $\mathbf{B}_{\rm A}=\delta B \mathbf{e}_\phi$, and the wave electric field has purely poloidal components. The inductive
(ideal) electric field amplitude  is
$\mathbf{E}_{\rm A}=-(\boldsymbol{\Omega}/c\times\mathbf{R})\times\mathbf{B}_{\rm dip}$ for purely toroidal shear velocities $v^\phi=\Omega R_\ast/c$. Inserting the dipolar field components from Equation~(\ref{eq:dipoleinit}) allows us to compute the wave amplitude $\delta E^2=\mathbf{E}_{\rm A}\cdot\mathbf{E}_{\rm A}$. We approximate the corresponding magnetic field amplitude by assuming $\delta B_A\approx\delta E_A$.}: 
\begin{align}
    b_0\equiv\frac{\delta B_A}{B_{\rm dip}}(\epsilon_0,\theta_s)\approx \left(\frac{\epsilon_0 R_\ast}{c}\right)\sin\theta_s,
\end{align}
where $\theta_s$ is an injection colatitude \citep[following closely the setups by][]{yuan2021,Chen2022,Bernardi2024}. The energy of an Alfvénic perturbation $\delta B$ propagating purely along the magnetic field is conserved within a dipolar flux tube. If the wave does not experience significant shear, the product $\delta B^2 A$ between the pulse energy density and flux tube area $A$ is approximately constant. Along such a tube, the conservation of magnetic flux translates into the relation $B_1/B_2=A_2/A_1$, where the subscripts represent two different locations along the flux tube. The strength of the magnetic perturbation scales as
\begin{align}
    \delta B\propto B_{\rm dip}^{1/2}\propto r^{-3/2}.
\end{align}
Thus, the relative strength follows $b\propto B_{\rm dip}^{-1/2}\propto r^{3/2}$. We define the critical nonlinearity radius $\tilde{r}$ as the radius where $b=1$, such that
\begin{align}
    \tilde{r} = b_0^{-2/3}R_\ast.
    \label{eq:linearityestimate}
\end{align}
In any of the previous expressions, we label field lines anchored to the star by their radius of equatorial crossing, $R_{\rm eq}=R_{*}/\sin^2\theta_s$. The relative strength of the magnetic perturbation at the equator is $b_{\rm eq}=b_0\sin^{-3}\theta_s$. FMS waves propagate like vacuum electrodynamic modes in FFE. As they can cross magnetic field lines, their dynamics on a background magnetic field often resemble concentric waves like those spreading on the surface of calm water. In the axisymmetric dipole magnetic field, the polarization of FMS waves requires toroidal wave electric fields $\delta\mathbf{E}_F=E \mathbf{e}_\phi$ \citep[see detailed reviews by, e.g., ][]{Li2019}. As FMS waves have magnetic fields parallel or antiparallel to the background magnetic field, they can develop zones where the electric field nearly (or, even completely) dominates. In these `electric zones' the wave electric field becomes comparable to the total magnetic field \citep{Beloborodov2023,Chen2022b,Levinson2022}. The location where electric zones develop on the rapidly decaying dipole field ($B\propto r^{-3}$) depends on the colatitude, the injection location, and the injection amplitude. For an injection of an FMS wave at the stellar surface, we estimate the nonlinearity radius:
\begin{align}
    \frac{R_{\rm nl}(\theta)}{R_\ast}\approx \left(\frac{4-3\sin^2\theta}{2b_0\sin\theta}\right)^{1/2}.
    \label{eq:Rnl}
\end{align}
This estimate follows the derivation by \citet[][Equation~110]{Beloborodov2023}. Once electric zones develop, they can drive plasma heating as well as shock formation \citep{Beloborodov2023,Chen2022b}. We discuss the observational implications for the scenario of magnetospheric Alfvén wave collisions in Section~\ref{sec:discussion}.

\subsection{Mode conversion at curved field lines}
\label{sec:modeconvert}

In this subsection, we consider electric and magnetic fields $\mathbf{E}=\delta\mathbf{E}$ and $\mathbf{B}=\mathbf{B}_0+\delta\mathbf{B}$. Here, $\delta\mathbf{E}$ and $\delta\mathbf{B}$ are perturbations to a background field $\mathbf{B}_0$ that is not necessarily dipolar. The force-free current on a stationary (i.e., $\nabla\times\mathbf{B}_0=0$) background expressed in these variables reads (up to second order-perturbations) as
\begin{align}
\begin{split}
    &\mathbf{J}_{\rm FFE} \\
     &=\left(\nabla\cdot\mathbf{E}\right)\frac{\mathbf{E}\times\mathbf{B}}{B^2}+\frac{\mathbf{B}\cdot\left(\nabla\times\mathbf{B}\right)-\mathbf{E}\cdot\left(\nabla\times\mathbf{E}\right)}{B^2}\mathbf{B}\\
    &=\left(\nabla\cdot\delta\mathbf{E}\right)\frac{\delta\mathbf{E}\times\mathbf{B}_0}{B_0^2}+\frac{\delta\mathbf{B}\cdot\left(\nabla\times\delta\mathbf{B}\right)}{B_0^2}\mathbf{B}_0\\
    &\qquad +\frac{\mathbf{B}_0\cdot\left(\nabla\times\delta\mathbf{B}\right)}{B_0^2}\delta\mathbf{B}-\frac{\delta\mathbf{E}\cdot\left(\nabla\times\delta\mathbf{E}\right)}{B_0^2}\mathbf{B}_0\\
    &\qquad -\frac{2\mathbf{B}_0\cdot\left(\nabla\times\delta\mathbf{B}\right)}{B_0^4}\left(\delta\mathbf{B}\cdot\mathbf{B}_0\right)\mathbf{B}_0\\
    &\qquad+\frac{\mathbf{B}_0\cdot\left(\nabla\times\delta\mathbf{B}\right)}{B_0^2}\mathbf{B}_0.\label{eq:pertcurrent}
\end{split}
\end{align}
An Alfvén wave with $\omega=\pm ck_\parallel$ in an axisymmetric dipole magnetosphere with $\mathbf{B}_0=\mathbf{B}_{\rm dip}$ as described by Equation~(\ref{eq:dipoleinit}) has a wavevector in the $\left(r,\theta\right)$--plane \citep[a comprehensive analysis of such waves can be found in][]{chen2024}. The electric perturbation is also in the $\left(r,\theta\right)$--plane and perpendicular to $\mathbf{B}_{\rm dip}$. The magnetic perturbation is along the $\phi$ direction. Their amplitudes are related by $\delta B=-\delta E$. Analysis of wave-like perturbations shows that if the current defined in Equation~(\ref{eq:pertcurrent}) has a second-order contribution along the $\phi$ direction \citep[due to curvature effects such that the first three terms do not cancel out as for straight field lines, see][]{chen2024}, it has the dependence
\begin{align}
    \frac{J_{\phi, \rm FFE}}{B_{\rm dip}}\propto i k_\perp\left(\frac{\delta B}{B_{\rm dip}}\right)^2.
\end{align}
If the Alfvén wave cannot compensate for this current by matching a suitable $\nabla\times\delta\mathbf{B}$, a toroidal electric perturbation can be induced. Such perturbations can seed FMS propagating outward. The strength of such a fast wave (F) injection can be roughly constrained by analyzing Equation~(\ref{eq:MaxIII}) in the limit of no $\nabla\times\delta\mathbf{B}$ compensation. It then depends on the seed Alfvén wave (A) amplitude (with respect to the background field) and its variation $k_\perp$ perpendicular to the magnetic field. We can further approximate
\begin{align}
    \frac{\delta E_F}{B_{\rm dip}}\propto\frac{4\pi}{c}\frac{k_\perp}{k_\parallel}\left(\frac{\delta B_A}{B_{\rm dip}}\right)^2\equiv\frac{4\pi}{c}\xi \left(\frac{\delta B_A}{B_{\rm dip}}\right),\label{eq:fastcouple}
\end{align}
where $\xi = (k_\perp/k_\parallel)(\delta B_A/B_{\rm dip})$ is a measure of Alfvén wave (non)linearity. We analyze the dipolar magnetic fields along the two coordinates
\begin{align}
    p = \frac{r}{\sin^2\theta}=R_{\rm eq}\qquad\qquad q=\frac{\cos\theta}{r^2}\equiv\frac{\chi}{r^2},
\end{align}
where $\left(r,\theta\right)$ are the radius and colatitude of the corresponding spherical coordinate system \citep{Swisdak2006,Chen2022}. The $p$-coordinate labels each field line by its radius on the magnetic equator ($\chi=0$); $q$ varies along the length $s\left(p,\chi\right)$ of the field line. Alfvén waves traveling along the dipole field develop significant shear with a rapid increase of $\xi$. The shear evolution is driven by the gradient
\begin{align}
    \begin{split}
        \nabla_\perp s=\frac{\sqrt{1+3\chi^2}}{\sin^3\theta}&\left[F\left(\chi_0\right)-F\left(\chi\right)\phantom{\frac{\sqrt{1+3\chi_0^2}}{2\chi_0}}\right.\\
        +&\left.\frac{\sqrt{1+3\chi_0^2}}{2\chi_0}\frac{R_\ast}{p}-2\frac{\chi\sin^2\theta}{\sqrt{1+3\chi^2}}\right].
    \end{split}
    \label{eq:kshear}
\end{align}
We use the stellar radius $R_\ast$, the footpoint colatitude $\sin^2\theta_0=R_\ast/R_{\rm eq}$, and $F$ as defined by \citet{Chen2022}. Equation~(\ref{eq:kshear}) predicts the evolution of the ratio $k_\perp/k_\parallel$ between perpendicular and parallel scales of the wave. Integrating Equation~(\ref{eq:kshear}) and scaling $\delta B_A/B_{\rm dip}$ along a magnetic field line, we can use Equation~(\ref{eq:fastcouple}) to estimate the strength of coupling between Alfvén waves propagating along curved field lines and the injected fast modes (Figure~\ref{fig:COUPLING}):
\begin{align}
    \frac{\delta E_F}{B_{\rm dip}}\propto \frac{4\pi}{c}\left(\nabla_\perp s\right)\sin^6\theta\;b_{\rm eq}^2.
    \label{eq:couplingestimate}
\end{align}
The coupling is strongest at colatitudes in the hemisphere opposite to the injection site. Specifically, Figure~\ref{fig:COUPLING} has a maximum beyond the equator ($\theta\approx 110^\circ$). The coupling strength scales with the wave-driven magnetic stress $b_{\rm eq}^2$. We note that the fast wave properties outlined in this section are only rough estimates of the expected trends. We ultimately rely on fully fledged FFE simulations (Section~\ref{sec:simulations}) to solve the nonlinear field evolution and consider the complex field line geometry.

\subsection{Nonlinear Alfvén wave interaction}
\label{sec:nonlinear}

In addition to mode conversion at curved field lines (Section~\ref{sec:modeconvert}), fast waves can be generated by resonant three-wave interaction between Alfvén waves.
For resonant interactions, the kinetic condition of momentum and energy conservation must be satisfied:
\begin{eqnarray}
    \mathbf{k}_1+\mathbf{k}_2 &=&\mathbf{k},\label{eq:res1}\\
    \omega_1+\omega_2 &=&\omega.
\end{eqnarray}
Subscripts $1$ and $2$ denote the incoming Alfvén waves; unsubscripted quantities refer to the outgoing FMS wave. The coefficient that determines the growth rate of secondary modes of the three-wave interaction is given, for example, in \citet[][Appendix~A]{Li2019}
\begin{align}
\begin{split}
    C_{\mathcal{A}\mathcal{A}\mathcal{F}'}&\propto\frac{\omega}{B_0\sqrt{\omega \omega_1\omega_2}} \frac{\delta B_1\delta B_2}{k_{\perp}k_{1\perp}k_{2\perp}}\left[\left( \omega_1\omega_2 - c^2k_{1\parallel}k_{2\parallel}\right)\right.\\ &\times\left.(2k_{1\perp}^2k_{2\perp}^2 + \left(k_{1\perp}^2+k_{2\perp}^2\right)\mathbf{k}_{1\perp}\cdot \mathbf{
	k}_{2\perp} )\right]\, .
\end{split}
\label{eq:interactioncoefficient}
\end{align}
Here, $\delta B_1$ and $\delta B_2$ are amplitudes for the interacting Alfvén waves, and $B_0$ is the background magnetic field strength. For two identical waves with reversed polarization $\boldsymbol{k}_{1\perp}=-\boldsymbol{k}_{2\perp}$, this coefficient will vanish; there will be no fast wave produced from the three-wave interaction. With similar arguments than in our derivation of Equation~(\ref{eq:couplingestimate}), one can infer a significant spatial dependence of the interaction coefficient: off-equatorial interactions or asymmetric wave injection scenarios could change -- likely decrease -- FMS wave coupling. The solution space to the kinetic condition (Equation~\ref{eq:res1}) depends on the resolved dimensions and the allowed wavevectors $\mathbf{k}$. Capturing space in three dimensions will allow more orientations of the propagating waves and, thus, more resonant channels than in two dimensions. Therefore, considering three-dimensional setups of the vast parameter space in future work may lead to different conversion dynamics \citep[including relativistic turbulence; see, e.g.,][]{Li2019,Ripperda2021,TenBarge2021}. 
We discuss the implications and limitations of dimensionality in more detail in Section~\ref{sec:limitations}.

\section{Simulations}
\label{sec:simulations}

In this work, we model the propagation and interaction of magnetic perturbations in axisymmetric magnetar magnetospheres in FFE simulations. We employ a high-order FFE method with optimized hyperbolic/parabolic cleaning parameters \citep{Mahlmann2020b,Mahlmann2020c,Mahlmann2021} that vastly benefits from the \textsc{Carpet} driver \citep{Goodale2002a,Schnetter2004}. Its extension to spherical coordinates \citep{Mewes2018,Mewes2020} is supported by the infrastructure of the \textsc{Einstein Toolkit} \citep{Loeffler2012,yosef_zlochower_2022_6588641}\footnote{\url{http://www.einsteintoolkit.org}}. 

\subsection{Setup}
\label{sec:setup}

\begin{figure*}
\centering
  \includegraphics[width=0.975\linewidth]{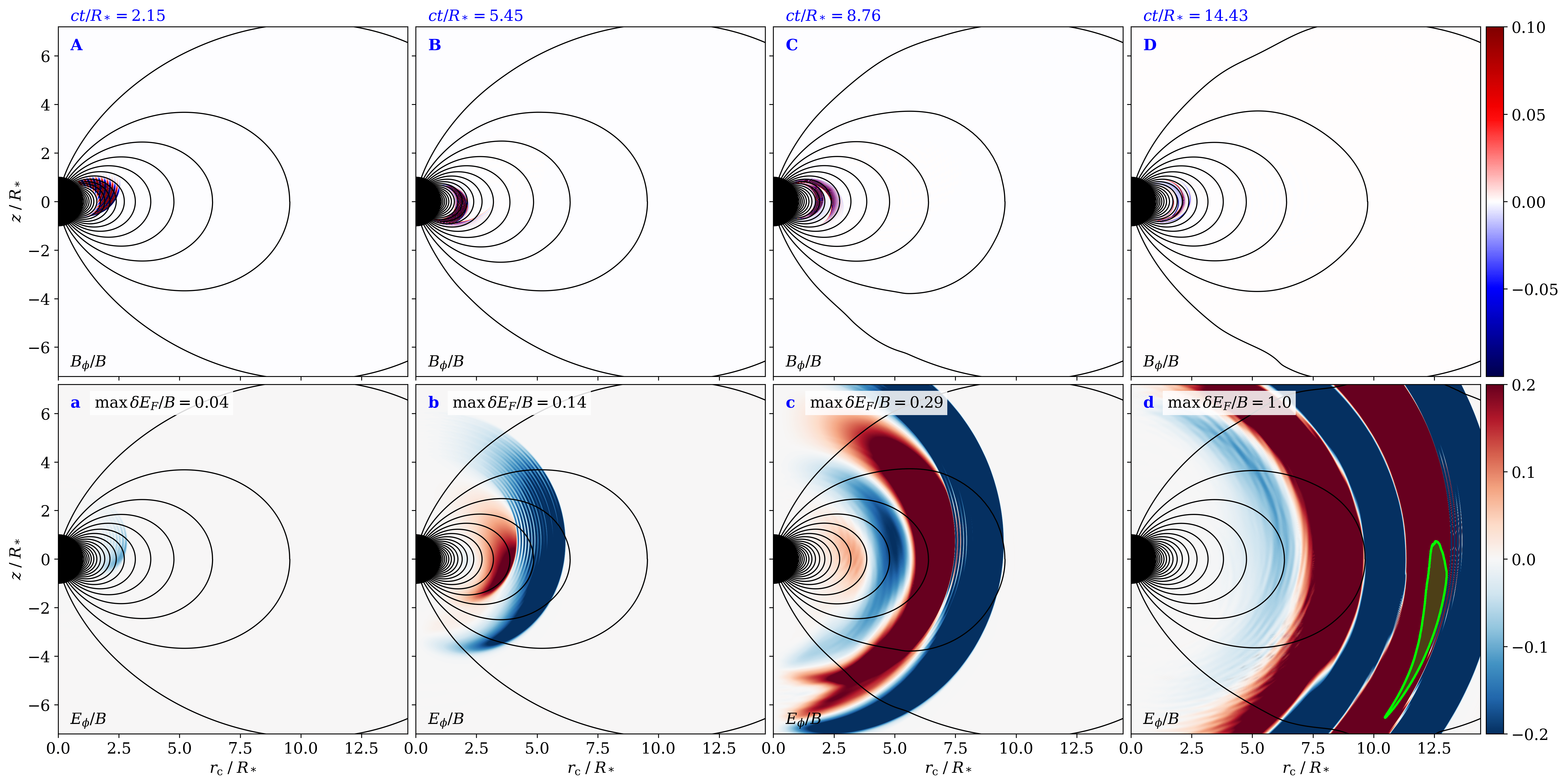}\\
  \vspace{-5pt}
  \flushleft
  \hspace{0pt}\includegraphics[width=0.935\linewidth]{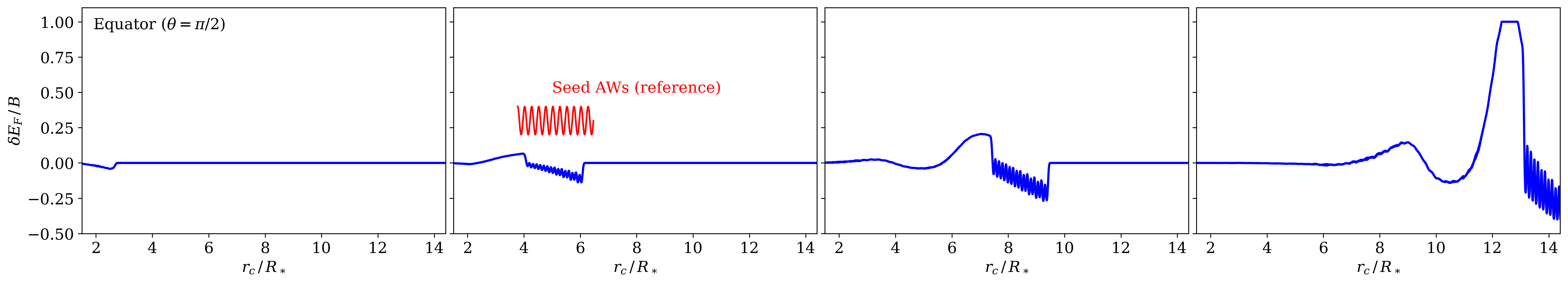}
  \caption{Injection and mode conversion of an Alfvén wave envelope with $b_{\rm eq}=0.55$, $cT/\lambda=8.0$ and $\lambda=R_\ast/4$, located at $\theta_s=45^\circ$ with width $\Delta\theta=5^\circ$, simulated with $N_r(N_{R_\ast})\times N_\theta = 19968 (512)\times 1600$. The horizontal axis shows the cylindrical radius $r_c=r\sin\theta$. The initially linear Alfvénic perturbation propagates along magnetic field lines, as visualized in the top row. Interaction with the curved magnetic field lines converts part of the Alfvén wave energy into fast magnetosonic modes propagating across field lines, see middle row. Alfvén waves are strongly sheared due to the different lengths of magnetic field lines (panel C) and bounce in the inner magnetosphere after reflection from the stellar surface (panels C/D). The fast wave amplitude relative to the magnetic field becomes significant as the generated pulses propagate outward (panel d, green contours denote $E>B$). We indicate the relative strength of the FMS wave amplitude in the middle panels. Bottom panels: Outline of the FMS wave electric field $\delta E_F/B=r_cE^\phi/B$ along the equator ($\theta=\pi/2$) at the time of the upper panels. We display the waveform of the seed Alfvén waves in red color, as a reference (not true to scale). We provide an animated version of this figure as supplementary material \citep{SupplementaryMediaA}. It shows the evolution of the magnetic fields (left animation panel) and electric fields (right animation panel) during times $ct/R_\ast=0$ to $15.75$; the real-time duration of the animation is $7\,{\rm s}$.}
  \label{fig:a1bounce}
\end{figure*}


All our simulations initialize a dipolar magnetic field anchored to a perfect conductor of radius $R_\ast$ following Equation~(\ref{eq:dipoleinit}). We use spherical coordinates with axisymmetry \citep[similar to][on the same infrastructure]{Mahlmann2021}. The computational domain encloses the area $(r,\theta)\in\left[R_\ast, 40R_\ast\right]\times\left[0,\pi\right]$ with uniform spacing. We vary the resolution by changing the number of grid points in each direction, with $N_r\in\left\{4992,9984,19968,39936\right\}$ and $N_\theta=\left[400,800,1600,3200\right]$. We also measure the radial resolution by the number of grid points per stellar radius $N_{R_\ast}\in\left\{128,256,512,1024\right\}$. Beyond $r/R_\ast = 40$, the radial grid spacing increases by a factor $a=1.001$ in each cell, effectively allowing a logarithmic spacing up to very large radii. We integrate this system with seventh-order accurate spatial reconstruction \cite[MP7;][]{Suresh1997} and $\text{CFL}\approx 0.2$.

To drive the desired magnetospheric dynamics, we initialize an Alfvén pulse in the magnetosphere by mimicking the toroidal displacement of the magnetar crust. Such displacements are localized in regions centered at colatitudes $\theta_{\rm s}$, with total widths $\Delta\theta$ \citep[as in][Equation 13]{Parfrey2013}. The maximum displacement is located in the center of these spots. Electric field boundary conditions at the surface of the central object drive the toroidal field line motion:
\begin{align}
\begin{split}
    \delta E^r_{\rm s} = -\epsilon r^2 \sin\theta B^\theta\qquad
    \delta E^\theta_{\rm s} = \epsilon \sin\theta B^r.
\end{split}
\end{align}
Here, the magnetic field is that of a dipole as denoted in Equation~(\ref{eq:dipoleinit}). We set the amplitude $\epsilon$ of the displacement to vary in time according to the waveform
\begin{align}
    \epsilon = \epsilon_0 \sin\left(\frac{2\pi n}{T}\left[t-t_{\rm min}\right]\right)
\end{align}
for intervals $t_{\rm min}\leq t\leq t_{\rm min}+T$. The factor $n$ denotes the number of waves with wavelength $\lambda/c=T$ injected into the domain. 

\subsection{Wave trains `bounce' in the inner magnetosphere}
\label{sec:trains}

In this section, we consider an Alfvén wave packet with a long envelope that can drive continuous wave interaction in the inner magnetosphere. The boundary conditions inject Alfvén waves with $cT/\lambda=8.0$ and $\lambda=R_\ast/4$, for a fixed injection colatitude $\theta_{\rm s}=45^\circ$ and width $\Delta\theta = 5^\circ$. We vary the wave amplitude as well as the numerical resolution of the computational domain. 

\begin{figure}
\centering
  \includegraphics[width=0.975\linewidth]{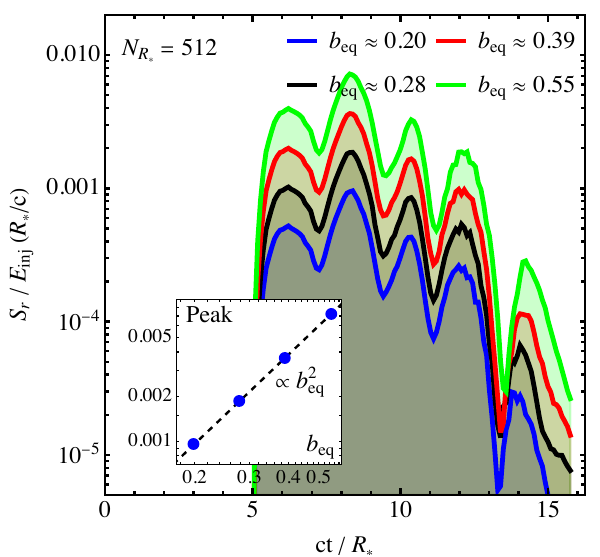}
  \caption{Outgoing FMS wave luminosity injected by long-envelope Alfvén waves (see Figure~\ref{fig:a1bounce}) measured at $r/R_\ast=5$ and normalized to the injected Alfvén wave energy $E_{\rm inj}$. We vary the injection amplitude for a fixed numerical resolution. The inset plot shows the peak luminosity as a function of the wave amplitude $b_{\rm eq}$, following a trend $\propto b_{\rm eq}^2$.}
  \label{fig:PNT_AMP}
\end{figure}

We first fix the injection amplitude such that $b_{\rm eq}\approx 0.2$. Figure~\ref{fig:a1bounce} \citep[and animation in the supplementary material;][]{SupplementaryMediaA} shows an Alfvén wave envelope propagating through the inner magnetosphere and building up significant shear (panel A). The sheared Alfvén waves inject FMS waves (panel a) as described in Section~\ref{sec:enhancement}. Once the Alfvén waves reach the stellar surface of the opposite hemisphere, they bounce off it and propagate back (panel B/C). FMS waves with amplitudes of opposite signs are produced
upon each Alfvén wave crossing (panels b/c/d). The bottom panels of Figure~\ref{fig:a1bounce} display short-wavelength modulations of the generated FMS waves. Fourier analysis of the modulation shows that their wavelength is half of the wavelength of the injected Alfvén waves (see Appendix~\ref{app:fft}). The high-frequency variations of seed Alfvén waves only imprint on the outgoing FMS waves with twice the Alfvén wave frequency during the first magnetospheric crossing. Later FMS waves have a much longer wavelength comparable to the timescale $T_{\rm bounce}$ of Alfvén waves propagating through the magnetosphere. For the chosen injection colatitude of $\theta_s=45^\circ$, field lines have a length of $L\approx 3.5 R_\ast$. The bottom panels of Figure~\ref{fig:a1bounce} illustrate the growth of FMS wave amplitude relative to the background magnetic field. Relatively small FMS production amplitudes ($\delta E_F/B\lesssim 0.05$) can grow to significant relative field strengths (with total field strengths $E\approx B$) within a few tens of stellar radii (see Figure~\ref{fig:a1bounce}, panel d, and Equation~\ref{eq:Rnl}). We note that, in this example, the electric zones develop only for the second half of the first 
generated FMS wave packet, the preceding high-frequency component of opposite $\delta E_F$ polarity propagates unobstructed. In Figure~\ref{fig:PNT_AMP}, we analyze how Alfvén waves with increasing amplitudes $b_{\rm eq}$ translate into a wave Poynting flux $\mathbf{S}=\delta\mathbf{E}\times\delta\mathbf{B}$ associated to outgoing FMS waves. Integrated over a spherical surface, we obtain the luminosity $L_r$. The peak luminosity measured at $r/R_\ast=5$ scales with the square of the Alfvén wave amplitude. Equation~(\ref{eq:fastcouple}) predicts this enhancement with the magnetic stress carried by the AW. The peaks of the luminosity are separated by $ct/R_\ast\approx 2$ (see further discussion below). 

\begin{figure}
\centering
  \includegraphics[width=0.975\linewidth]{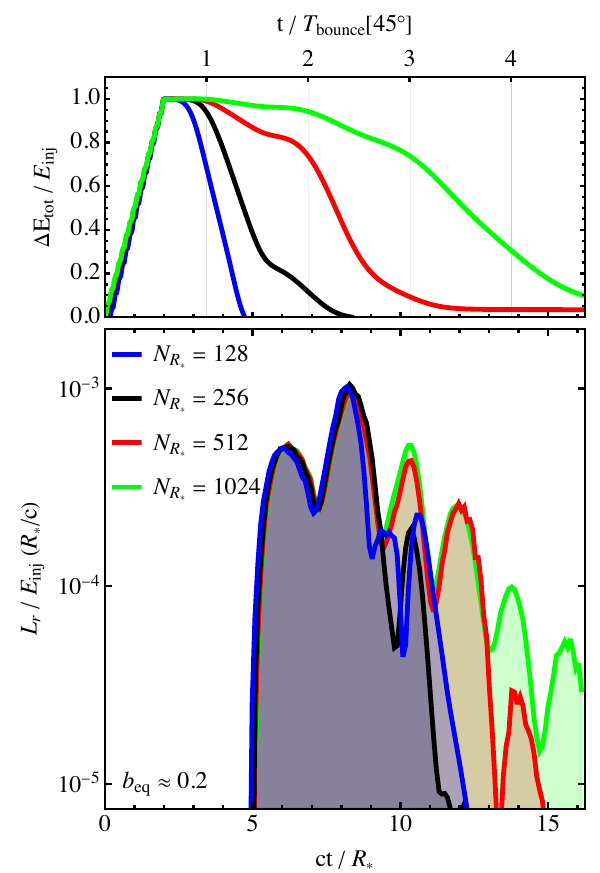}
  \caption{Resolution requirements for late-stage FMS wave generation. The bottom panel is as Figure~\ref{fig:PNT_AMP} but for a fixed injection amplitude ($b_{\rm eq}\approx 0.2$) and varying numerical resolution. The top panel shows the excess magnetospheric energy $\Delta E_{\rm tot}=E_{\rm tot}-E_{\rm dip}$, normalized to the injected Alfvén wave energy $E_{\rm inj}$. For an injection colatitude of $\theta_s=45^\circ$, field lines have a length of $L\approx 3.5 R_\ast$; we denote the light-crossing time of this length as $T_{\rm bounce}$.}
  \label{fig:PNT_RES}
\end{figure}

The strong shear that Alfvén waves experience (especially after several magnetospheric crossings) makes them susceptible to grid diffusion. Therefore, it is instructive to probe the numerical convergence of the FMS wave injection. Figure~\ref{fig:PNT_RES} (bottom panel) shows the luminosity of outgoing FMS waves measured at $r/R_\ast =5$ for varying resolutions $N_{R_\ast}\in\left\{128,256,512,1024\right\}$. Higher resolutions capture the sheared waves significantly better. The top panel of Figure~\ref{fig:PNT_RES} tracks the evolution of the excess energy $\Delta E_{\rm tot}=E_{\rm tot}-E_{\rm dip}$, which corresponds to the energy injected in Alfvén waves initially, and to the combined energy in Alfvén waves and FMS waves at later times $t\gtrsim T$. For $N_{R_\ast}=128$, the entire injected Alfvén wave energy dissipates right after the first magnetospheric `bounce' of duration $T_{\rm bounce}$ (here, this scale denotes the time it takes a seed Alfvén wave to cross the magnetosphere along a dipolar field line). Higher resolutions delay the decay of excess energy to later times. Hence, the decay is driven by numerical diffusion as the energy cascades down to scales of the grid resolution. The highest resolution of $N_{R_\ast}=1024$ allows for several magnetospheric bounces with mild diffusion at the grid scale. Our results show that all the chosen resolutions capture the initial phase of FMS wave production equally well. Only after the second magnetospheric crossing and $T_{\rm bounce}\gtrsim 2$, the FMS wave energy significantly deviates for the lower resolution setups. The total efficiency of FMS wave conversion ranges between $0.23\%$ ($N_{R_\ast}=128$), $0.23\%$ ($N_{R_\ast}=256$), $0.28\%$ ($N_{R_\ast}=512$), and $0.30\%$ ($N_{R_\ast}=1024$). For these parameters, the FMS energy varies by $30\%$, attributed purely to the numerical resolution capturing (or not) the strong shear of Alfvén waves in the decaying tail of FMS wave pulses. In Appendix~\ref{app:tailfit}, we estimate that $0.16\%$ of the injected Alfvén wave energy (around \emph{half} of the entire energy transferred to FMS waves) is converted to FMS waves in the tail after $ct/R_\ast \approx 8.43$. Focusing only on this period emphasizes the damping effect of grid diffusion for late times\footnote{We use this comparison to an approximate `true' amount of energy stored in the tail (Appendix~\ref{app:tailfit}) to estimate the order of convergence achieved with our method in resolving the decaying FMS wave tail. For $N_{R_\ast}\gtrsim 256$, we empirically find a convergence order of $1.27$. This is significantly below the theoretical convergence order of our method \citep[seventh order for smooth regions, cf.][]{Mahlmann2020c}. The break-down of convergence order is a strong indication of the difficulties in resolving the ongoing force-free wave interactions in the inner magnetar magnetosphere.}: the highest probed resolution ($N_{R_\ast}=1024$) recovers around $90\%$ of the tail energy $\tilde{E}_{\rm tail}$, the lowest probed resolution ($N_{R_\ast}=128$) only recovers $38\%$ and fails to resolve the tail modulation with $T_{\rm bounce}$.

\subsection{Prompt generation of FMS waves}
\label{sec:enhancement}

\begin{figure*}
\centering
  \includegraphics[width=0.975\linewidth]{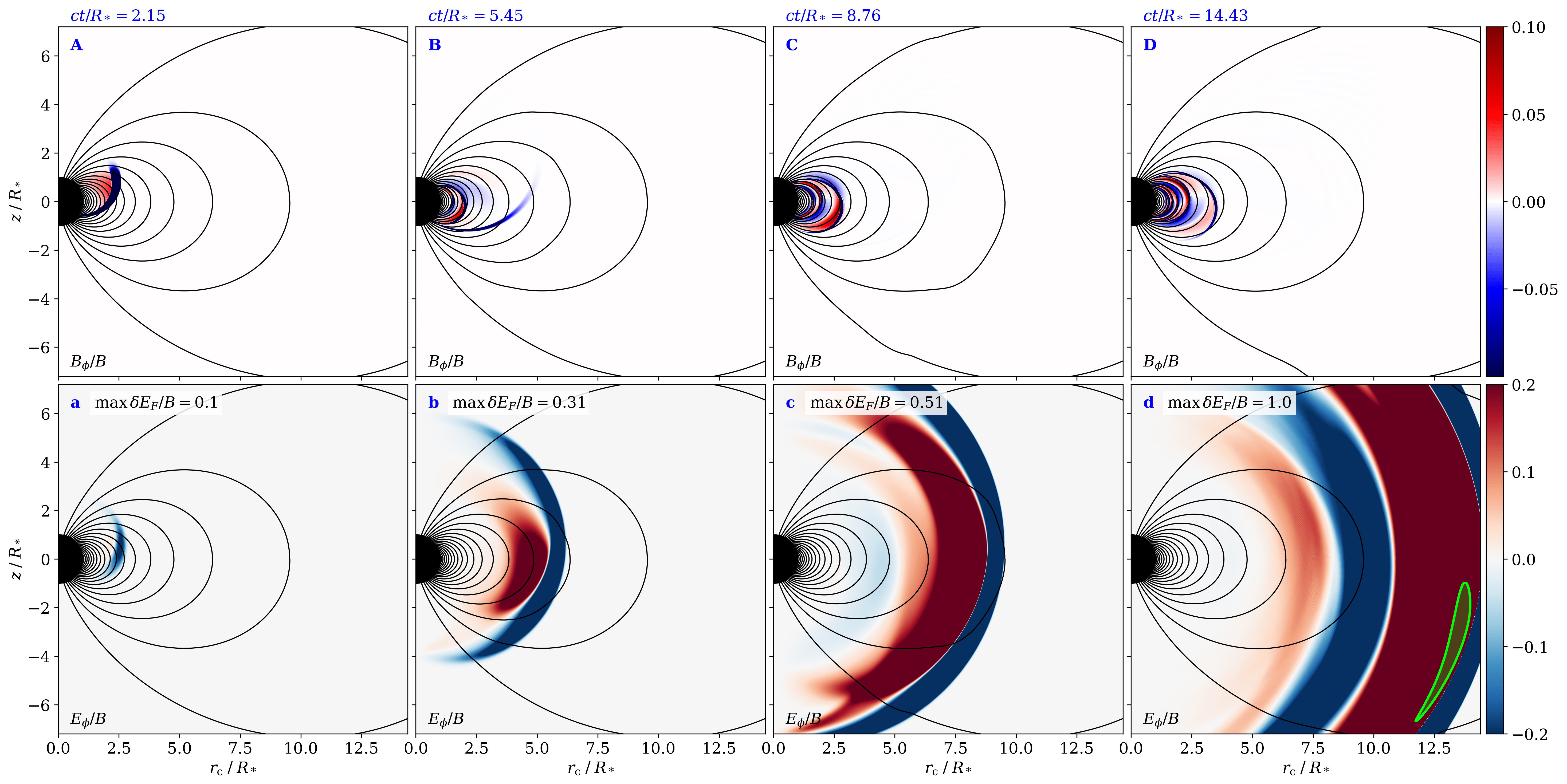}
  \caption{As Figure~\ref{fig:a1bounce} but for a single Alfvén wave injected with $b_{\rm eq}=0.39$ and half a wavelength ($cT/\lambda=0.5$, $\lambda=R_\ast/2$) at $\theta_s=45^\circ$ with width $\Delta\theta=20^\circ$ and resolution $N_r(N_{R_\ast})\times N_\theta = 19968 (512)\times 1600$. The fast wave amplitude relative to the magnetic field can develop extended electrically dominated zones with $E>B$ (green contours, panel d). We provide an animated version of this figure as supplementary material \citep{SupplementaryMediaB}. It shows the evolution of the magnetic fields (left animation panel) and electric fields (right animation panel) during times $ct/R_\ast=0$ to $16.79$; the real-time duration of the animation is $8\,{\rm s}$.}
  \label{fig:a1history}
\end{figure*}

\begin{figure*}
\centering
  \includegraphics[width=0.975\linewidth]{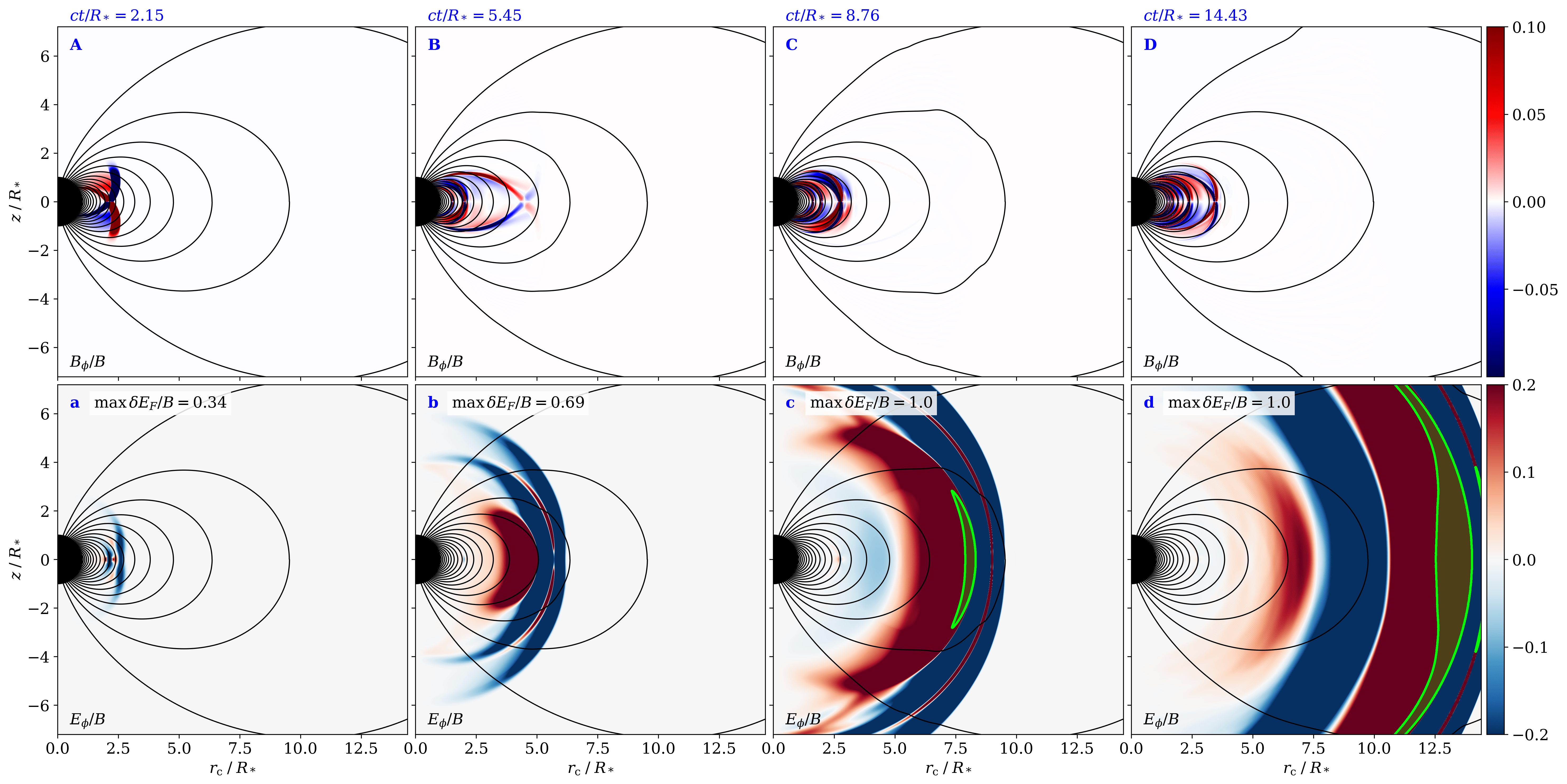}
  \caption{As Figure~\ref{fig:a1history}, but for two Alfvén waves launched symmetrically and out-of-phase ($\varphi = \pi$) in both hemispheres. The Alfvén waves interact in the equatorial region (panels B/C/D) and alter the fast wave generation. We provide an animated version of this figure as supplementary material \citep{SupplementaryMediaC}. It shows the evolution of the magnetic fields (left animation panel) and electric fields (right animation panel) during times $ct/R_\ast=0$ to $16.79$; the real-time duration of the animation is $8\,{\rm s}$.}
  \label{fig:a2history}
\end{figure*}

\begin{figure}
\centering
  \includegraphics[width=0.99\linewidth]{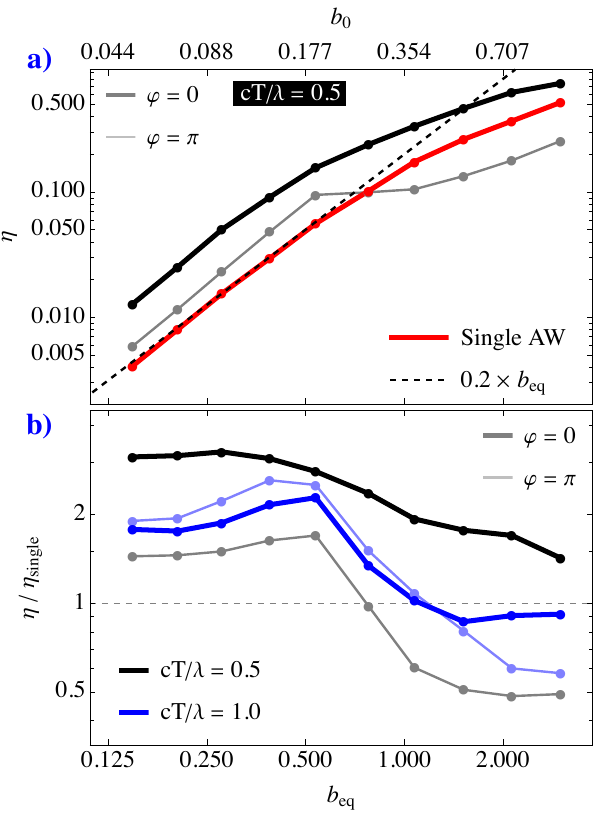}
  \caption{Conversion efficiency of injected Alfvén wave energy to outgoing FMS wave energy using $N_r(N_{R_\ast})\times N_\theta = 4992 (128)\times 400$. Panel a) shows the conversion efficiency for seed Alfvén waves packets of $cT/\lambda=0.5$. The red line denotes a single Alfvén wave injected in the northern hemisphere, black and gray lines show conversion efficiencies for seed Alfvén waves in both the northern and the southern hemispheres. Panel b) displays conversion efficiencies of Alfvén wave interactions normalized to the efficiency of a single wave. We show different wavelengths of the injected Alfvén wave (black and blue color) and in-phase versus out-of-phase wave injection in the two hemispheres (thick and thin lines, respectively). 
  }
  \label{fig:EFFICIENCY}
\end{figure}

In this section, we explore how the interaction of counter-propagating Alfvén waves affects the FMS wave generation during their first emission peak. Single Alfvén waves propagating along curved magnetic field lines can seed FMS waves \citep[see Section~\ref{sec:modeconvert}, and][]{yuan2021,chen2024}. The efficiency of FMS wave production
can change due to nonlinear Alfvén wave interactions (see Section~\ref{sec:wavedynamics}). 
The first magnetospheric crossing of the seed Alfvén waves generates a prompt peak of FMS waves that carries a significant fraction of the total FMS energy. We use the smallest numerical resolution ($N_{R_\ast}=128$) that resolves this phase and conduct a broad parameter scan. We ensure sufficiently resolved seed Alfvén waves by choosing large injection regions with $\Delta\theta=20^\circ$. As a reference for direct comparison, Figure~\ref{fig:a1history} shows the propagation of a single seed Alfvén wave through the magnetosphere \citep[see animation in the supplementary material,][]{SupplementaryMediaB}. Moving along the magnetic field direction, the wave experiences significant shear due to the different lengths of the supporting magnetic field lines (Figure~\ref{fig:a1history}, top panels). With increasing shear and relative amplitude of the seed Alfvén waves, FMS waves emerge from the inner magnetosphere (bottom panels). The relative amplitude of FMS waves grows as they propagate across the decaying dipole magnetic field (see Equation~\ref{eq:linearityestimate}). Electric zones with $E\approx B$ develop due to the propagation effects discussed in Section~\ref{sec:wavedynamics} (Figure~\ref{fig:a1history}, green contours). 

To measure the effect of nonlinear wave interactions we inject counter-propagating Alfvén waves in both hemispheres. In this scenario (Figure~\ref{fig:a2history}), Alfvén waves propagate through the magnetosphere symmetrically and interact in the equatorial region \citep[top panels, see also animation in the supplementary material, out-of-phase:][in-phase: \citealt{SupplementaryMediaD}]{SupplementaryMediaC}. FMS waves are produced and become nonlinear in the extended magnetosphere (bottom panels, green contours). To quantify the conversion efficiency, we measure the total Alfvén wave energy injected into the interaction region, $E_{\rm inj}$, by evaluating the energy flux through a sphere close to the stellar surface. We then compare it to the total energy leaving the interaction region as FMS waves, $E_{\rm FMS}$ (measured at some distance from the star). The conversion efficiency is $\eta\equiv E_{\rm FMS}/E_{\rm inj}$. Figure~\ref{fig:EFFICIENCY} panel a) displays conversion efficiencies for different combinations of seed Alfvén waves with wavelength $\lambda=0.5R_\ast$ injected for half a wavelength, $cT/\lambda = 0.5$. The reference case of a single Alfvén wave injected in the northern hemisphere with the same properties is shown in red color and can be compared directly to the scaling found by \citet[][dashed line]{yuan2021}. We probe phase differences of $\varphi=0$ and $\varphi=\pi$ for the interacting half-wavelength modes: the conversion efficiency for in-phase Alfvén waves is significantly enhanced. 

In Figure~\ref{fig:EFFICIENCY} panel b), we compute $\eta$ for interacting waves and compare them to those of single waves $\eta_{\rm single}$ varying both envelope size  $cT/\lambda\in \left\{0.5,1.0\right\}$ and phase shift 
$\varphi\in\left\{0,\pi\right\}$. When $\eta/\eta_{\rm single}\lesssim 1$, FMS wave generation is reduced compared to the single wave scenario, while for $\eta/\eta_{\rm single}\gtrsim 1$ the FMS wave production is enhanced. Significant enhancement of the FMS wave generation efficiency happens for short envelopes with $cT/\lambda=0.5$ that are in-phase ($\varphi=0$). The enhancement is reduced but not fully suppressed for out-of-phase waves ($\varphi=\pi$), broadly consistent with the analysis of the three-wave interaction coefficient (Section~\ref{sec:nonlinear}). Even though the seed Alfvén waves are counter-propagating, the conversion coefficient does not vanish due to the shear experienced by waves propagating along dipolar field lines. For increasing wave amplitudes (i.e., larger $b_{\rm eq}$), the conversion efficiency for all configurations decreases compared to that of a single wave. Then, the single wave propagation in a nonuniform (curved) background magnetic field becomes highly nonlinear and dominates over Alfvén wave interaction for FMS wave generation. The creation of electric zones with rapid dissipation at the collision center \citep{li2021f} explains the significant suppression of FMS wave production for high-amplitude seed waves with magnetic perturbations that cancel out upon superposition (possible regardless of phase shift for waves of at least one full wavelength $cT/\lambda=1.0$).

\subsection{Repeated interaction of counter-propagating Alfvén waves}
\label{sec:counterprop}

\begin{figure}
\centering
  \includegraphics[width=0.975\linewidth]{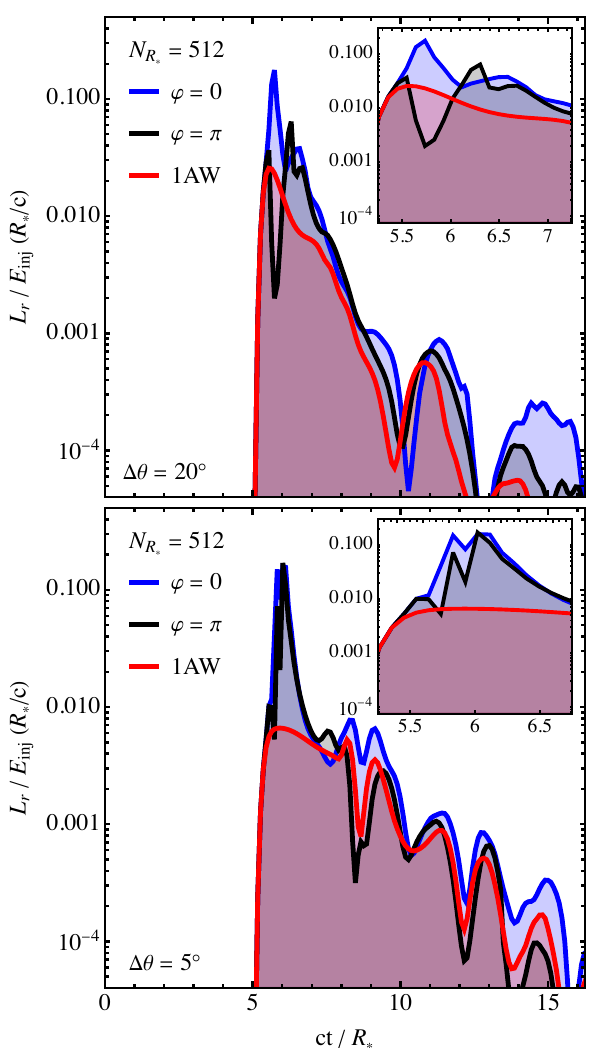}
  \caption{Outgoing FMS wave luminosity injected by short-envelope Alfvén waves. We compare counter-propagating Alfvén waves ($b_{\rm eq} \approx 0.39$, $cT/\lambda=0.5$, $\lambda=R_\ast/2$, $\theta_s=45^\circ$) of different phases (black/blue lines) and varying sizes of the injection region (top: $\Delta\theta= 20^\circ$; bottom: $\Delta\theta= 5^\circ$) with the FMS wave injected by a single Alfvén wave (red lines). The inset shows a zoom into the initial interaction period, where the phase difference induces different FMS wave luminosities.}
  \label{fig:PNT_INTERACT}
\end{figure}

In this section, we study the enhanced FMS wave generation for short-envelope seed Alfvén waves in more detail. With high numerical resolution ($N_{R_\ast}=512$), we revisit Alfvén waves injected with $cT/\lambda=0.5$, $b_{\rm eq}=0.39$ (see Section~\ref{sec:enhancement}), for different angular extensions $\Delta\theta=5^\circ$ and $\Delta\theta=20^\circ$. Again, we compare the FMS wave generation during magnetospheric crossings of a single Alfvén wave with the dynamics of two Alfvén waves injected in both hemispheres with phases $\varphi\in\left\{0,\pi\right\}$. Figure~\ref{fig:PNT_INTERACT} shows the time-resolved luminosity carried by escaping FMS waves for injection regions with $\Delta\theta = 5^\circ$ (bottom panel) and $\Delta\theta = 20^\circ$ (top panel). Alfvén waves without phase difference are the most efficient generators of outgoing FMS waves. Independent of $\Delta\theta$, the energy transported by FMS waves increases by an order of magnitude during a short initial spike. Out-of-phase waves have a reduced conversion efficiency that depends on the width $\Delta\theta$ of the injection site. Waves with large $\Delta\theta$ show a significant decrease in conversion efficiency: approximately two orders of magnitude compared to the phase-aligned case. For the chosen seed waves, we find efficiencies $\eta_{\varphi = 0}/\eta_{\rm single}\approx 3.1$, and $\eta_{\varphi = \pi}/\eta_{\rm single}\approx 1.6$, hence, a decrease of approximately $50\%$. Waves with smaller $\Delta\theta$ have a less severe decrease of FMS generation. For the chosen seed waves, we find efficiencies $\eta_{\varphi = 0}/\eta_{\rm single}\approx 4.5$, and $\eta_{\varphi = \pi}/\eta_{\rm single}\approx 3.2$, hence, a decrease of approximately $30\%$. Such enhancements were suggested in Section~\ref{sec:enhancement} and here we confirm them for selected cases with higher numerical resolution.

\begin{figure}
\centering
  \includegraphics[width=0.975\linewidth]{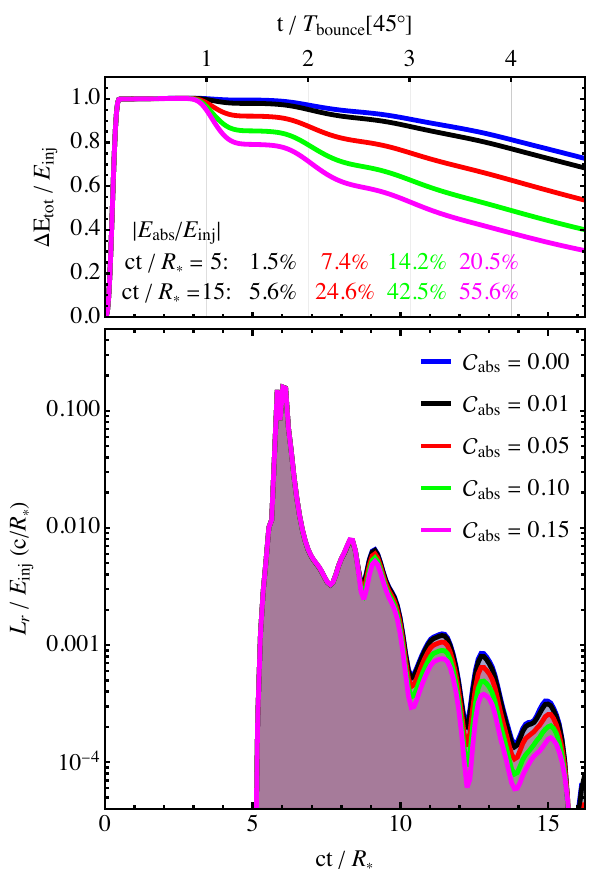}
  \caption{Magnetospheric energy balance with Alfvén wave absorption at the stellar surface. We vary the absorption coefficient $\mathcal{C}_{\rm abs}$ (see Equation~\ref{eq:cabs}). The bottom panel shows the outgoing FMS wave luminosity. The total energy of generated FMS waves decreases at later times. Top panel: as in Figure~\ref{fig:PNT_RES}, numbers show the amount of absorbed energy.}
  \label{fig:PNT_ABS}
\end{figure}
Besides the strong shear Alfvén waves experience when propagating through the magnetosphere, they can be damped plastically when interacting with the magnetar crust \citep[e.g.,][]{Li2015}. In the last set of tests, we evaluate the dynamics of two Alfvén waves injected in opposite hemispheres that experience an energy loss every time they interact with the stellar surface boundary. We adapt the perfectly conducting boundary used throughout this work \citep[see details in][]{Mahlmann2021,Mahlmann2023} by including an absorbing layer close to the star. Following \citet[][equation 53]{Parfrey2012pheadra}, we add damping terms that drive the field components of the Alfvén wave polarization ($B^\phi$, $E^r$, $E^\theta$) to the dipole field. We vary the damping rate $\mathcal{C}_{\rm abs}$ for a smooth damping profile active within $r<1.1R_\ast$. In practice, we drive the absorption by adding source terms of the form 
\begin{align}
    \mathcal{S}_{\rm abs}[B^\phi]= -\mathcal{C}_{\rm abs}B^\phi
    \label{eq:cabs}
\end{align}
to the finite volume integration of the respective field components. For the same seed wave properties as above ($b_{\rm eq} \approx 0.39$, $cT/\lambda=0.5$, $\lambda=R_\ast/2$, $\theta_s=45^\circ$, $\Delta\theta=5^\circ$), we increase the absorption strength $\mathcal{C}_{\rm abs}$. Figure~\ref{fig:PNT_ABS} shows how the magnetospheric excess energy evolves in time (top panel), and how the damping of Alfvén waves impacts the outgoing FMS waves (bottom panel). For the explored values of $\mathcal{C}_{\rm abs}$, we compute the relative change of generated FMS wave energy as follows:
\begin{align}
\begin{split}
    \frac{\Delta E_{\rm FMS}}{E_{\rm inj}}&\equiv \frac{E_{\rm FMS}-E_{\rm FMS}\left(\mathcal{C}_{\rm abs}=0\right)}{E_{\rm inj}}.
\end{split}
\end{align}
We then approximate the wave energy absorbed by the boundary layer as the difference $E_{\rm abs}\equiv\Delta E_{\rm tot}-\Delta E_{\rm tot}\left(\mathcal{C}_{\rm abs}=0\right)$, neglecting the small contribution of FMS waves to the total excess energy. The plastic damping of Alfvén waves at the stellar boundary only has a small effect on the outgoing FMS waves. While the absorption strength $\mathcal{C}_{\rm abs}$ significantly reduces the total excess energy in the magnetosphere (see Figure~\ref{fig:PNT_ABS}, top panel), the outgoing FMS wave luminosity decreases only slightly at late times. Specifically, the FMS wave efficiency drops by $0.23\%$ ($\mathcal{C}_{\rm abs}=0.01$), $1.13\%$ ($\mathcal{C}_{\rm abs}=0.05$), $2.06\%$ ($\mathcal{C}_{\rm abs}=0.10$), and $2.83\%$ ($\mathcal{C}_{\rm abs}=0.15$). The initial FMS pulse -- emitted before the first boundary interaction of the seed Alfvén wave -- does not experience additional losses. For $t/T_{\rm bounce}[45^\circ]\gtrsim 2$, the excess energy (mainly stored in Alfvén waves) slowly decays with time. The rate of this late-stage decay is similar for all the probed absorption coefficients $\mathcal{C}_{\rm abs}$. This coincidence of dissipation rates is likely a combination of two effects. First, for the employed resolution ($N_{R_\ast}=512$) sheared Alfvén waves could be affected by grid diffusion for $t/T_{\rm bounce}[45^\circ]\gtrsim 2$. The chosen resolution $N_{R_\ast}$ is a trade-off between computational feasibility and an accurate representation of the relevant physics. Second, as the Alfvén waves become increasingly sheared, they could lose their wave character ($\omega=ck_\parallel\approx 0$, see also Equation~\ref{eq:kshear}) and remain in the inner magnetosphere as an extended shear. Such a state would be less affected by the boundary absorption acting only very close to the star. Figure~\ref{fig:PNT_ABS} (top panel) displays the fraction of absorbed energy at different times. We discuss the implications of these wave dynamics in the following sections.

\section{Discussion}
\label{sec:discussion}

This paper analyzes the conversion efficiency between Alfvén waves confined to the inner magnetar magnetosphere and outgoing fast magnetosonic waves that can escape the (inner) magnetosphere. Such mode conversions are an essential step to transport energy from close to the magnetar to far distances without significant disruptions of the magnetosphere itself \citep[like those discussed, e.g., by][]{Parfrey2013,Mahlmann2019,yuan2020,Carrasco_etal_2019_10.1093/mnrasl/slz016,Mahlmann2022,Sharma2023}. Faults and failures of the stressed magnetar crust likely inject Alfvén waves into the magnetosphere. \citet{Thompson2017} propose that crustal stresses can be released to the magnetosphere during short outbursts lasting several milliseconds. The development of crustal yielding in magnetars is still an area of active research \citep[e.g.,][]{Blaes1989,Thompson1996,Levin2012,Beloborodov2014,Bransgrove2020,Gourgouliatos2019,Lander2023}. General estimates \citep[see review in][Section~2]{yuan2020} suggest that crustal failures in a stressed volume $V$ process energies of around 
\begin{align}
    E_{\rm Quake}&\lesssim& V\mu_{s} \frac{s^2}{2} \approx 10^{44}\left(\frac{V}{10^{16}{\rm cm}^3}\right)\left(\frac{s}{0.1}\right)^2\,{\rm erg}.
\end{align}
Here, $\mu_{s}\approx 10^{30}\,{\rm erg}\,{\rm cm}^{-3}$ is the shear modulus of the deep crust and $s<0.1$ denotes the elastic strain. Alfvén waves emerging from the stressed regions during the quake duration $t_{\rm Quake}$ carry a fraction $f$ of $E_{\rm Quake}$ \citep{Bransgrove2020}. Their power is:
\begin{align}
    L_{\rm A}\approx 10^{44}\left(\frac{f}{0.01}\right)\left(\frac{E_{\rm Quake}}{10^{44}\,{\rm erg}}\right)\left(\frac{t_{\rm Quake}}{10\,{\rm ms}}\right)^{-1}{\rm erg}\,{\rm s}^{-1}.\label{eq:AWLUM}
\end{align}
Throughout Section~\ref{sec:simulations} we numerically evaluate conversion efficiencies $\eta$ for two different mode conversion mechanisms in axisymmetry: mode conversion at curved magnetic field lines and by nonlinear three-wave interactions. Typical conversion efficiencies range between $\eta\approx 1-10\%$ for low-amplitude waves; higher efficiencies can be achieved by counter-propagating interacting Alfvén waves and for large seed wave amplitudes (see Section~\ref{sec:enhancement}). Directly applying these efficiencies to Equation~(\ref{eq:AWLUM}) and assuming that most of the FMS wave energy is concentrated in the short initial pulse of the first magnetospheric crossing yields FMS wave luminosities of $L_{\rm FMS}\approx 10^{42}-10^{43}\,{\rm erg}/{\rm s}$. Alfvén waves can generate FMS waves of twice their frequency during the first crossing of the magnetosphere. Subsequent bounces produce more FMS waves during each crossing. These late-episode FMS waves gradually decrease in luminosity but extend the possible signal durations substantially. We describe the FMS wave tail generated by ongoing Alfvén wave activity in the inner magnetosphere by a fit function (see Appendix~\ref{app:tailfit}). Especially for the long-envelope wave trains discussed in Section~\ref{sec:trains}, a high-quality fit can recover characteristic scales of the problem, like the crossing time of the Alfvén waves through the magnetosphere ($T_b\approx T_{\rm bounce}$). We find that different injection scenarios -- like the interacting waves analyzed in Section~\ref{sec:counterprop} -- show decaying tails with less regularity and are harder to approximate by the proposed fit function. The difference in the decaying tails likely implies that late-stage dynamics are affected by the initial perturbations, especially the number of seed Alfvén wave envelopes and their length. 

Plastic damping of Alfvén waves during their interaction with the magnetar surface \citep[e.g.,][]{Li2015} can speed up the dissipation of the initially injected magnetospheric excess energy. Realistic absorption mechanisms likely remove $\gtrsim10\%$ of Alfvén wave energy in each surface interaction, where most of the Alfvén wave energy is quickly absorbed and dissipated inside the star. However, the impact of crustal damping on the luminosity of outgoing FMS waves is small (see Figure~\ref{fig:PNT_ABS}). The produced FMS waves gain most of their energy during the first bounce of the injected Alfvén waves, as they interact around the equatorial plane. Hence, the most significant production of FMS wave energy occurs before the plastic damping acts on the reflected Alfvén waves. Electric zones of $E\approx B$ can naturally develop in parts of FMS waves leaving the inner magnetosphere with polarizations such that the wave magnetic field is directed opposite to the stellar magnetic field \citep{Beloborodov2023,Chen2022b,Mahlmann2022}. In the following section, we discuss the implications of these findings for FRBs.

\subsection{Implications for Fast Radio Bursts}

FRBs are bright millisecond-duration flashes of radio waves in the gigahertz band \citep[$0.1-10$\,GHz,][]{Chime2019b}, with at least one observational association to a galactic magnetar \citep[SGR 1935+2154][]{Chime2020}. Between many possible generation mechanisms \citep[reviews, e.g., in][]{Lyubarsky2021,Petroff2022,Zhang2023}, two scenarios for radio wave generation in the outer magnetar magnetosphere have been modeled consistently at relevant plasma scales: shock-mediated FRB generation at magnetized shocks by the synchrotron maser mechanism \citep[e.g.,][]{Ghisellini2016,Plotnikov2019,Metzger2019} or mode conversion \citep{Thompson2022} on the one hand, and reconnection-mediated radio wave injection by in the compressed magnetospheric current sheet on the other hand \citep{Lyubarsky2021,Mahlmann2022,wang2023}.

Reconnection-mediated FRB generation models rely on a long-wavelength FMS pulse (roughly millisecond duration) propagating to the outer magnetosphere up to the magnetar light cylinder. Mode conversion of Alfvén waves at curved field lines can generate such long pulses in two ways. First, magnetospheric crossings of long-wavelength\footnote{The terms `short wavelength' and `long wavelength' can be interpreted with respect to different reference scales. For example, the wavelength $\lambda$ can be compared to the length $L$ of the field line along which a seed Alfvén wave propagates. Alternatively, $\lambda$ could be compared to the scale of variation of the background magnetic field $L_B\sim B_{r,\rm dip}/|\text{d}B_{r,\rm dip}/\text{d}r|\sim r/3$. In Section~\ref{sec:trains}, seed Alfvén waves have \emph{short} wavelengths of $\lambda_0=R_\ast/4\lesssim L_B<L$. The first train of generated FMS waves has shorter wavelengths of $\lambda_1=R_\ast/8<L_B<L$. Later FMS modes have \emph{long} wavelengths $\lambda_2\approx L\approx 3.5R_\ast>L_B$. These scales are separated by a factor $\lambda_2/\lambda_1\approx 30$.} seed Alfvén waves generate FMS waves (Section~\ref{sec:counterprop}) with conversion efficiencies $\eta$ as evaluated in Section~\ref{sec:enhancement}. Second, later bounces of seed Alfvén waves result in FMS waves with modulations at scales of the Afvén-crossing time of a tube of dipolar field lines anchored to the magnetar surface at both ends (Section~\ref{sec:trains}). The bottom panels of Figure~\ref{fig:a1bounce} lucidly illustrate this production of late-phase long-wavelength modes. The first magnetospheric crossing generates high-frequency FMS waves with wavelengths prescribed by the seed Alfvén waves. The subsequent bounce induces a long-wavelength FMS pulse with a wavelength defined by the light-crossing time of the Alfvén waves in the inner magnetosphere. For the chosen injection colatitude of $\theta_s=45^\circ$, field lines have a length of $L\approx 3.5 R_\ast$. The light-crossing time for dipolar field lines is 
\begin{align}
    T_{\rm bounce}\approx 0.12\left(\frac{L}{3.5R_\ast}\right) \left(\frac{R_\ast}{10^6{\rm cm}}\right)\,{\rm ms}.
\end{align}
Therefore, bounces in the inner magnetosphere ($\theta_s=45^\circ$) could provide long-wavelength FMS pulses of duration $T_{\rm bounce}\approx 0.1\,{\rm ms}$. This long-wavelength time scale can be compared to the observational properties of FRBs: their total duration ($\lesssim 1\,{\rm ms}$), their short duration substructures \citep[e.g.,][]{PastorMarazuela2023,Snelders2023}, and their quasi-periodic behavior \citep[e.g., with hundreds of millisecond periods, see][]{Andersen2022}. Alfvén wave activity in the inner magnetosphere could generate FMS waves with variability on these timee scales. Especially the pulse width would be compatible with the low-frequency FMS pulse required for reconnection-mediated FRB generation in the outer magnetosphere \citep[see further review of feedback mechanisms by][]{Mahlmann2024}. Stressing field lines closer to poles will increase the bounce time. For instance, it approaches typical FRB durations of $T_{\rm bounce}\approx 1\,{\rm ms}$ as required by reconnection-mediated models for $\theta_s=17.5^\circ$. From the bottom panel of Figure~\ref{fig:a1bounce} one can infer directly that for the chosen seed Alfvén waves ($b_{\rm eq}=0.2$), the injected FMS waves have amplitudes of $\delta E_F/B=r_c E^\phi/B\lesssim 0.05$ at $r/R_\ast\approx 2$. On the rapidly decaying dipole field, relative FMS wave amplitudes scale as $\delta B/B_{\rm dip}\propto r^2$ \citep[e.g.,][]{Chen2022b} and develop very large amplitudes compared to the magnetic background at the magnetar light cylinder. FMS wave amplitudes as simulated in this work would surpass the compression required by reconnection-mediated FRB models by several orders of magnitude \citep[see][Equation~28]{Mahlmann2022}.

If the FMS wave magnetic fields are directed opposite to the magnetospheric fields, FMS waves can generate electric zones with $E\approx B$. Such zones are efficient sites of plasma heating \citep{Levinson2022} or particle acceleration in radiative shocks \citep{Beloborodov2023,Chen2022b}. For FMS waves generated close to the stellar surface, the radius at which electric zones develop for equatorial propagation is $R_{\rm nl}/R_\ast\approx(2b_0)^{-1/2}$. Very low injection amplitudes $b_0\approx 10^{-6}$ are sufficient to create nonlinear FMS waves with electric zones for $r/R_\ast\lesssim 10^3$. The wave dynamics discussed in Section~\ref{sec:trains} illustrate an important aspect of the electric-zone-paradigm: only those parts of the FMS waves that carry a magnetic field in the direction opposite to the background field develop $E\gtrsim B$ dynamics. Figure~\ref{fig:a1bounce} shows that the corrections only act on waves with one polarity of $\delta E_F$. The other polarity propagates unobstructed -- here carrying a high-frequency signal imprinted on the long-wavelength envelope. Besides instabilities in the inner magnetosphere \citep[see][]{Mahlmann2022,Mahlmann2024}, the mode conversion mechanism discussed in this paper can be an additional source of ubiquitous (sub-GHz) FMS waves leaving the inner magnetosphere and possibly developing magnetized shocks. Then, they are viable precursors for shock-mediated FRB generation in the outer magnetosphere. FMS waves that have different polarizations would still be able to provide the compression relevant for reconnection-mediated models.

\subsection{Limitations}
\label{sec:limitations}

This paper explores FFE simulations of wave propagation and dissipation in axisymmetric dipole fields to model some of the dynamics expected in highly magnetized magnetar magnetospheres. FFE simulations, though an efficient tool to model the propagation of magnetic energy in strong magnetic fields, have significant limitations. For the method employed in this work, we disclaim limitations exhaustively in \citet[][]{Mahlmann2020b,Mahlmann2020c,Mahlmann2022,Mahlmann2023}. Two aspects are highly relevant to the results presented above. First, our models include zones in which the electric field strength becomes comparable to the magnetic field strength ($E\approx B$). FFE methods enforce the magnetic dominance constraint $E<B$ of MHD by artificially dissipating the surplus electric field \citep[see detailed comments by][]{McKinney2006}. They do not capture the plasma kinetic feedback of electromagnetic dissipation accurately. We, therefore, omit statements about the dissipation of FMS energy in electric zones but refer the reader to the relevant works in the literature \citep{Levinson2022,Chen2022b,Beloborodov2023}. Future work will be conducted with methods that include the energy transfer between electromagnetic waves and magnetospheric plasma consistently (i.e., MHD or particle-in-cell). Second, Alfvén waves propagating along curved magnetic field lines develop significant shear that can quickly reach the limits of grid diffusion (Section~\ref{sec:trains}). In Figure~\ref{fig:PNT_RES}, we evaluate the resolution requirements for resolving the FMS wave injection during a significant number of Alfvén wave bounces in the inner magnetosphere. To resolve three magnetospheric crossings without significant diffusion, resolutions of $N_r(N_{R_\ast})\times N_\theta=19968(512)\times 1600$ are necessary, to resolve four to five bounces we need $N_r(N_{R_\ast})\times N_\theta=39936(1024)\times 3200$. These resolutions are very large even for axisymmetric configurations.

Constraining the magnetosphere to axisymmetry significantly limits the solution space for relativistic wave interactions \citep[e.g.,][]{Lyubarsky2018,TenBarge2021,Ripperda2021,Nattila2022,Golbraikh2023}. Giving accurate limits on conversion efficiencies and ultimately turbulent dissipation will require fully fledged three-dimensional models. The provided resolution limits can guide future magnetospheric simulations targeting wave interactions. The extreme requirements for resolving many wave interactions likely make comprehensive parameter scans prohibitive. Future work should therefore include analytic theory of magnetospheric waves and localized simulations with high resolution that accurately capture the dynamics in small parts of the magnetosphere.

\section{Conclusion}
\label{sec:conclusion}

This work reviews key aspects of relativistic plasma wave interactions in highly magnetized magnetar magnetospheres. We conduct FFE (see Section~\ref{sec:prelims}) simulations that follow the dynamics of Afvén waves injected into the magnetosphere by shearing the magnetar surface. These boundary conditions crudely mimic the possible stresses exerted on magnetic field lines by crustal failures. Alfvén waves traverse the magnetosphere along magnetic field lines; in dipolar geometries, they bounce back and forth between the hemispheres bounded by the highly conductive stellar surface. There are several mechanisms by which Alfvén waves generate FMS waves that can propagate outward, escape the inner magnetar magnetosphere, and drive energetic radiative outbursts. First, Alfvén waves propagating along curved field lines continuously inject FMS waves (Section~\ref{sec:modeconvert}). The coupling strength depends on the shear between adjacent field lines (i.e., the injection colatitude of seed Alfvén waves, see Figure~\ref{fig:COUPLING}) and the seed Alfvén wave amplitude. Second, counter-propagating Alfvén waves can interact to drive enhanced FMS wave injection (Section~\ref{sec:nonlinear}). The interaction of waves with short envelopes and no phase shift shows the largest conversion efficiency: here up to three times larger than for single Alfvén waves without interactions (Section~\ref{sec:enhancement}). Third, strongly sheared Alfvén waves can bounce in the inner magnetosphere. For example, long-envelope wave trains (Section~\ref{sec:trains}) generate high-frequency FMS waves during their first crossing. Subsequently, the remaining Alfvén waves drive low-frequency FMS wave packets. Initially with large amplitudes, their strength decreases with every bounce.

The injected FMS waves could be relevant for radiative phenomena observed from magnetars. FMS waves generated in the inner magnetosphere can be ubiquitous sources of electric zones, where the strengths of electric and magnetic fields become similar. These zones could drive shocks as one promising mechanism of FRB generation. Alfvén waves propagating through the inner magnetosphere (see Figure~\ref{fig:a1bounce}) could generate low-frequency magnetic perturbations and drive FRB generation in the outer magnetosphere. Understanding wave propagation and conversion in three dimensions will be key for explaining energy transport from the scale of crustal failures on the magnetar surface ($\sim 10^5\,{\rm cm}$) to the radiative scale (below $1\,{\rm cm}$). This work includes
important steps to a more complete theory of wave dynamics in the magnetar magnetosphere. 

\section*{Acknowledgements}

We thank the referee for feedback to our manuscript and the careful review of Section~\ref{sec:modeconvert}. The authors thank Andrei Beloborodov, Alex Chen, Amir Levinson, Alexander Philippov, Juan de Dios Rodríguez, Lorenzo Sironi, Anatoly Spitkovsky, Yajie Yuan, and Muni Zhou for useful discussions. We appreciate the support of the Department of Energy (DoE) Early Career Award DE-SC0023015 (PI: L. Sironi).
This work was supported by a grant from the Simons
Foundation (MP-SCMPS-00001470, PI: L. Sironi). This research was facilitated by the
Multimessenger Plasma Physics Center (MPPC), NSF
grant PHY-2206609. JFM acknowledges support from the National Science Foundation (NSF) under grant No. AST-1909458 (PI: A. Philippov) and AST-1814708 (PI: A. Spitkovsky).
MAA acknowledges support from grant PID2021-127495NB-I00, funded by MCIN/AEI/10.13039/501100011033 and by the European Union under `NextGenerationEU' as well as `ESF: Investing in your future'. Additionally, MAA acknowledges support from the Astrophysics and High Energy Physics program of the Generalitat Valenciana ASFAE/2022/026 funded by MCIN and the European Union `NextGenerationEU' (PRTR-C17.I1) as well as support from the Prometeo excellence program grant CIPROM/2022/13 funded by the Generalitat Valenciana.
JFM thanks the Department of Astronomy at Tsinghua University (Beijing, China) for their hospitality during a collaborative visit stimulating this research. The presented simulations were enabled by the \emph{Ginsburg} cluster (Columbia University Information Technology), the \emph{MareNostrum} supercomputer (Red Española de Supercomputación, AECT-2022-3-0010), and the \emph{Frontera} supercomputer \citep{Stanzione2020} at the Texas Advanced Computing Center (LRAC-AST21006). Frontera is made
possible by the NSF award
OAC-1818253. 

\bibliography{literature.bib}

\appendix

\section{Fit of the decaying FMS tail}
\label{app:tailfit}

\begin{figure}
\centering  \includegraphics[width=0.49\linewidth]{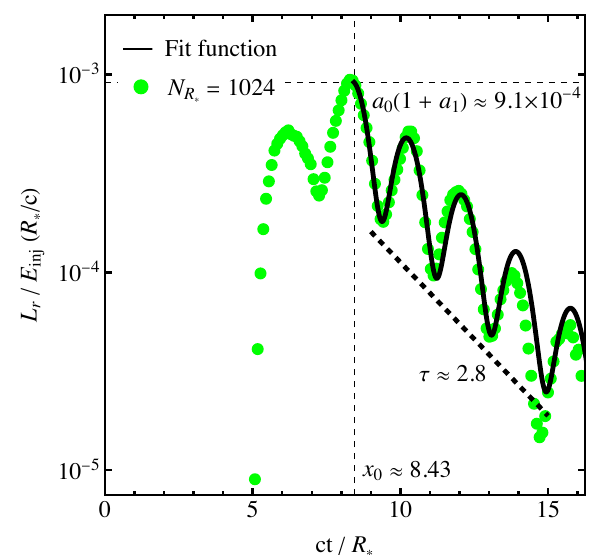}
  \caption{Approximation of the decaying tail of FMS pulses for the wave train discussed in Section~\ref{sec:trains} by a fit function (Equation~\ref{eq:fitfunction}). We indicate several fit parameters in the figure, including the decay onset time $x_0$ and the decay rate $\tau$ (following the dashed black line).}
  \label{fig:TailFit}
\end{figure}

In Section~\ref{sec:trains} we discuss the ongoing production of low-frequency FMS waves by Alfvén waves bouncing in the inner magnetospheres. At later times, Alfvén waves are strongly affected by grid diffusion due to their extreme shear accumulating during the propagation along the diverging dipole field lines. Figure~\ref{fig:PNT_RES} summarizes the extreme requirements on the numerical resolution for accurately capturing the decay of the outgoing FMS wave pulses. Here, we discuss how the expected luminosity function can be parametrized with an appropriate fit to approximate the total energy stored in the decaying FMS tail. We propose to analyze the expression:
\begin{align}
\Tilde{S}_r(x)/E_{\rm inj} (R_\ast/c)\approx a_0 \exp\left[-\frac{x-x_0}{\tau}\right]\times\left[\cos^2\left(2\pi\frac{x-x_0}{T_b} \right)+a_1\right],\qquad\text{where } x=\frac{ct}{R_*}.
\label{eq:fitfunction}
\end{align}
Here, the value of $a_0(1+a_1)$ corresponds to the luminosity flux at the time of the peak, $x_0$. The time scale $\tau$ is the characteristic decay rate, and $T_b$ denotes the modulation period. The parameter $a_1$ regularizes the modulation. This expression can be integrated to obtain the total energy $\tilde{E}_{\rm tail}/E_{\rm inj}$ stored in the outflowing FMS waves for times $x>x_0$:
\begin{align}
    \tilde{E}_{\rm tail}/E_{\rm inj}\equiv\int_{x_0}^\infty \text{d}x\,\Tilde{S}_r(x)/E_{\rm inj} (R_\ast/c)=a_0\tau\left(\frac{1}{2}+a_1+\frac{T_b^2}{T_b^2+16 \pi^2 \tau^2}\right).\label{eq:tailestimate}
\end{align}
Figure~\ref{fig:TailFit} shows a fit of Equation~(\ref{eq:fitfunction}) to the high-resolution ($N_{R_\ast}=1024$) data generated in Section~\ref{sec:trains}. Our numerical optimization yields the following fit parameters and standard errors: $a_0=(6.61\pm 0.19)\times 10^{-4}$, $a_1=(0.38 \pm 0.03)$, $x_0=(8.43\pm 0.01)$, $\tau=(2.80\pm 0.07)$, and $T_b=(3.71\pm 0.03)$. We can use these results to estimate the total energy $\tilde{E}_{\rm tail}/E_{\rm inj}\approx 0.16\%$ (Equation~\ref{eq:tailestimate}). Evaluating the fit function indicates that the modulation period $T_b$ of the decaying luminosity flux tail roughly corresponds to the Alfvén wave crossing time of the dipolar field line of the initial injection colatitude, where $cT_{\rm bounce}/R_\ast\approx 3.5$. We finally note that the amount of energy transferred to FMS waves during the first peak and the first half of the second peak is comparable to the energy in the tail, so that the total energy transferred from Alfvén waves to FMS waves is $\Delta E_{\rm FMS}/E_{\rm inj}\sim 0.3\%$.

\newpage

\section{Scales of the outgoing FMS waves}
\label{app:fft}

\begin{figure}
\centering  
\includegraphics[width=0.48\linewidth]{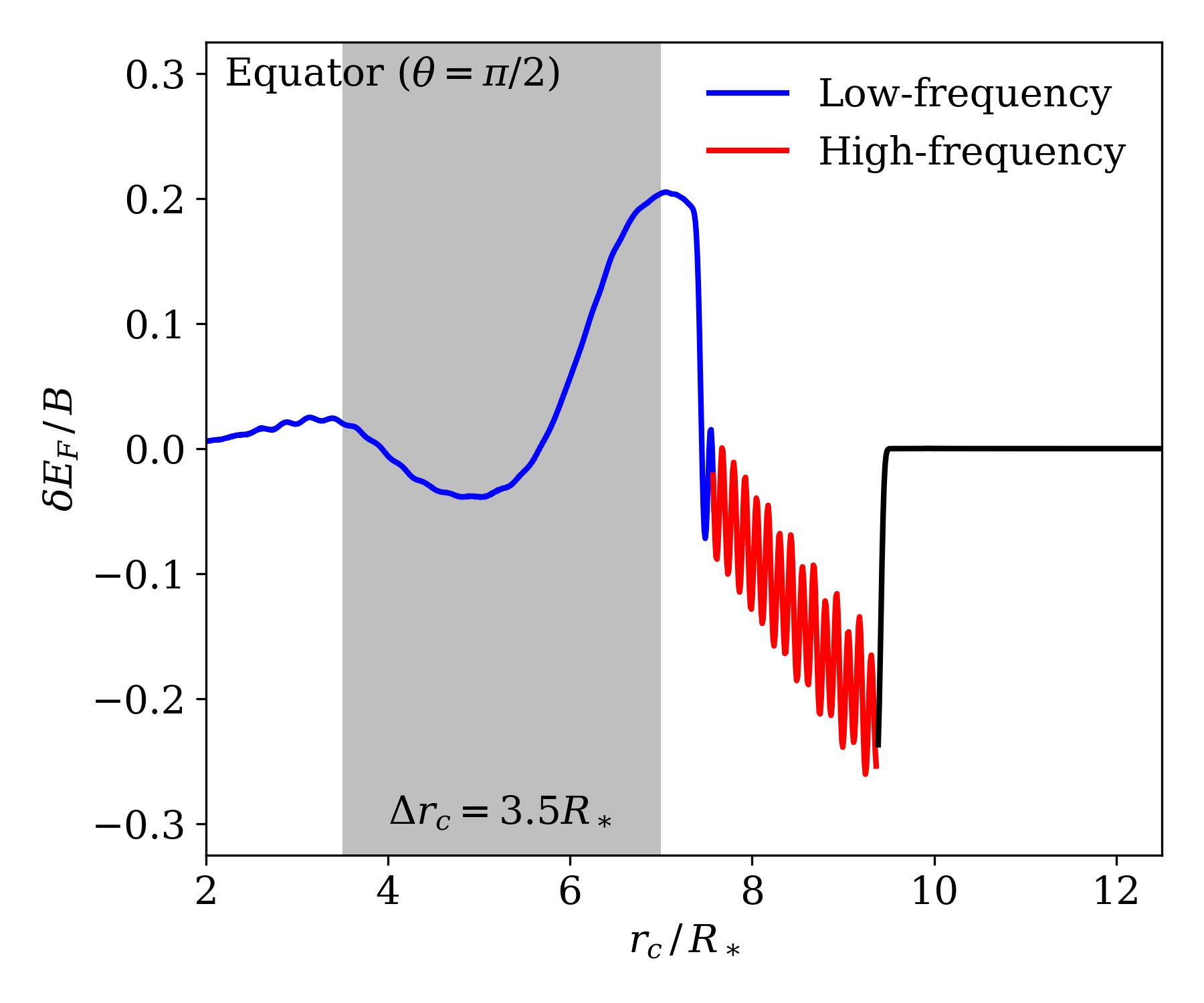}
\includegraphics[width=0.48\linewidth]{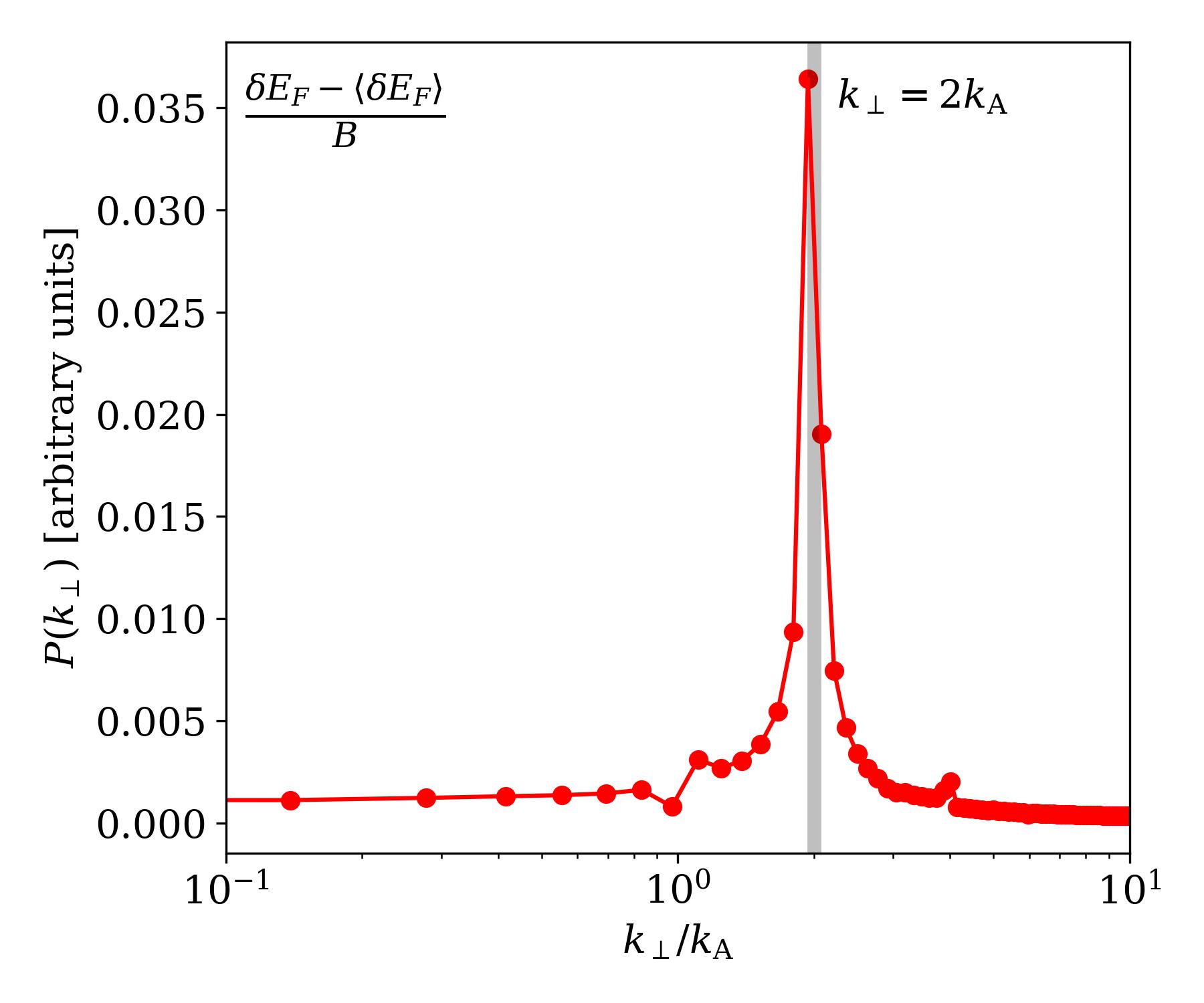}
  \caption{Analysis of the scales of FMS waves generated by the conditions outlined in Section~\ref{sec:trains} (see Figure~\ref{fig:a1bounce}). The left panel outlines the FMS wave along the equator with its high-frequency (red color) and low-frequency (blue color) components. We indicate the characteristic field line length for the propagation of seed Alfvén waves as a gray-shaded region. The right panel shows the power spectrum of the high-frequency FMS wave component (in arbitrary units). The average wave field is subtracted before the frequency analysis. The outgoing FMS waves have half the wavelength of the seed Alfvén waves.}
  \label{fig:fft}
\end{figure}

This section analyzes the properties of the FMS waves induced by the bouncing train of Alfvén waves discussed in Section~\ref{sec:trains}. For $ct/R_\ast = 8.76$ (Figure~\ref{fig:a1bounce}, third column), we assess the scales of the low-frequency and high-frequency wave components in Figure~\ref{fig:fft}. The high-frequency signal has a wavelength corresponding to half of the seed Alfvén wave wavelength, with a clear peak in the Fourier analysis shown in the right panel of Figure~\ref{fig:fft}. The variations of the low-frequency part correspond to the characteristic length $L\approx 3.5R_\ast$ of the field lines along which the seed Alfvén waves propagate.

\end{document}